\documentclass[journal]{IEEEtran}
\usepackage{psfrag}
\usepackage{subfigure}
\usepackage{url}
\usepackage{stfloats}
\usepackage{cite}
\usepackage{amsmath}
\usepackage{amssymb}
\usepackage{amsfonts}
\usepackage{graphicx}
\usepackage{fancybox}
\usepackage{color}
\usepackage{algorithm}
\usepackage{algorithmic}
\usepackage{multirow}
\usepackage{upgreek}
\usepackage{booktabs}
\usepackage{threeparttable}
\usepackage{latexsym}
\usepackage{bm,array}
\usepackage{graphicx}
\usepackage{float}
\usepackage{mathtools}
\usepackage{commath}
\usepackage{paracol}

\usepackage[version=4]{mhchem}

\usepackage{amsmath,lipsum}
\usepackage{cuted}%%\stripsep-3pt

\IEEEoverridecommandlockouts
\normalsize
\begin{document}

\title{Internet of Spacecraft for Multi-planetary Defense and Prosperity}
\author{
Yiming~Huo,~\IEEEmembership{Senior Member,~IEEE}
\thanks{Y. Huo is with University of Victoria, BC V8P 5C2, Canada (yhuo@ieee.org)}
}

\maketitle

\begin{abstract}
Recent years have seen unprecedentedly fast-growing prosperity in the commercial space industry. Several privately funded aerospace manufacturers, such as Space Exploration Technologies Corporation (SpaceX) and Blue Origin have innovated what we used to know about this capital-intense industry and gradually reshaped the future of human civilization. As private spaceflight and multi-planetary immigration gradually become realities from science fiction (sci-fi) and theory, both opportunities and challenges are presented. In this article, a review of the progress in space exploration and the underlying space technologies is firstly provided. For the next, a revisit and a prediction are paid and made to the K-Pg extinction event, the Chelyabinsk event, extra-terrestrialization, terraforming, planetary defense, including the emerging near-Earth object (NEO) observation and NEO impact avoidance technologies and strategies. Furthermore, a framework of the Solar Communication and Defense Networks (SCADN) with advanced algorithms and high efficacy is proposed to enable an internet of distributed deep-space sensing, communications, and defense to cope with disastrous incidents such as asteroid/comet impacts. Furthermore, the perspectives on the legislation, management, and supervision of founding the proposed SCADN are also discussed in depth.
\end{abstract}


\IEEEpeerreviewmaketitle
\newtheorem{mydef}{Definition}
\newtheorem{myLemma}{Lemma}
\newtheorem{theorem}{Theorem}
\newtheorem{Remark}{Remark}

\section{Introduction}
Since the first artificial Earth satellite, Sputnik 1, was sent into an elliptical low Earth orbit (LEO) by the Soviet Union (USSR) on October 4, 1957, humans have opened a gate to the space age. Another historical milestone was carved by National Aeronautics and Space Administration (NASA) Apollo 11, the first crewed spacecraft to land humans on another celestial body, Moon, on July 20, 1969 at 20:17 coordinated universal time (UTC). So far, more than thousands of spacecraft have been sent to outer space which is defined by the von Karman line at 100 km above Earth's mean sea level. By the generalized definition, a spacecraft is a vehicle or machine designed to fly in outer space, and it can be categorized into two major types: crewed and uncrewed. As of 2021, only three nations have flown crewed spacecraft, USSR/Russia, the USA, and China. The first crewed spacecraft, Vostok 1, was made by the Soviet Union and carried Soviet cosmonaut Yuri Gagarin into space in 1961, while the second crewed spacecraft named Freedom 7 was launched by United States also in 1961 and carried the first American astronaut Alan Shepard to an altitude of just over 187 kilometers (km). On the other hand, the uncrewed spacecraft includes Earth-orbit satellites, lunar probes, planetary probes, asteroid/comet probes, and other deep-space probes. More details about examples of both crewed and uncrewed spacecraft launched are presented in Table~\ref{table:spacecraft}.

\begin{figure} %[H]
\centering
\subfigure[]{\includegraphics[width=5 cm]{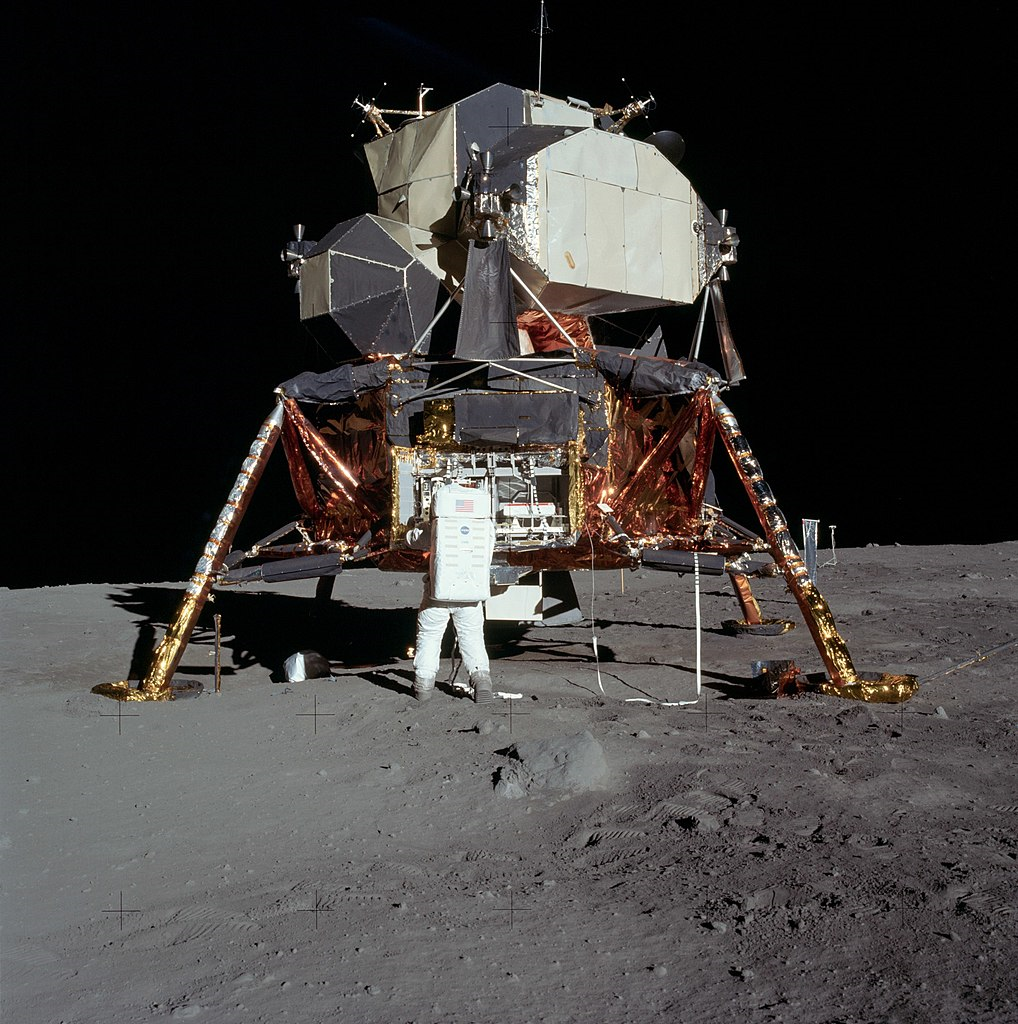}}
\subfigure[]{\includegraphics[width=7.9 cm]{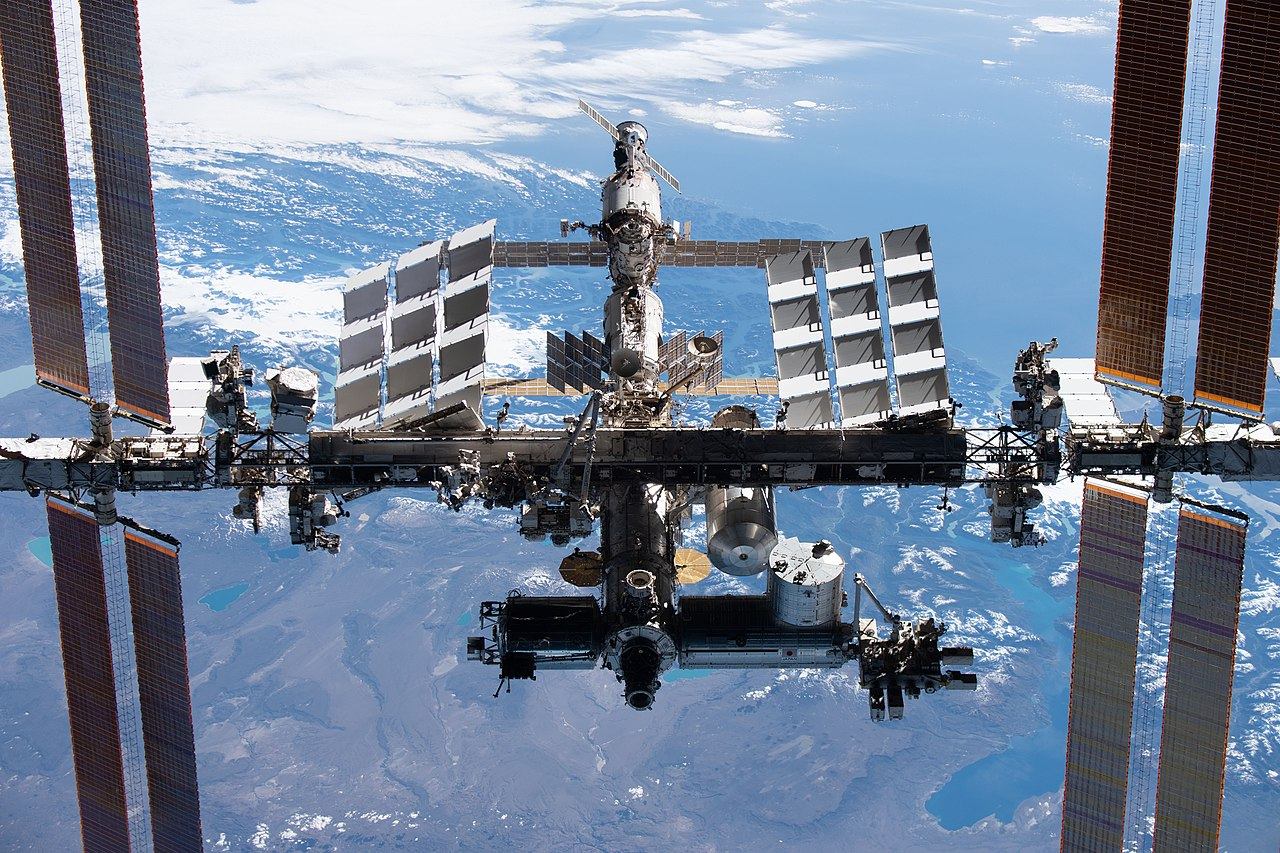}}
\caption{(\textbf{a}) Apollo 11 lunar module Eagle and astronaut Buzz Aldrin on the Tranquility Base of the Moon on July 21, 1969 \cite{ref-Armstrong}. (\textbf{b}) The view of ISS exterior and steelwork from the departing SpaceX Crew-2 spacecraft on Nov. 8, 2021 \cite{ref-ISS-2021}.} \label{MoonISS}
\end{figure}

\begin{table*}
\small
\caption{A summary of some crewed and uncrewed generalized spacecraft.\label{table:spacecraft}}
%%% \tablesize{} %% You can specify the fontsize here, e.g., \tablesize{\footnotesize}. If commented out \small will be used.
\newcommand{\tabincell}[2]{\begin{tabular}{@{}#1@{}}#2\end{tabular}}
\centering
\begin{threeparttable}
\begin{tabular}{|c|c|c|c|c|c|c|} \hline
%\toprule
\textbf{Category} & \textbf{Type}	& \textbf{Example}	& \textbf{Launch Year} & \textbf{Program Entity} & \textbf{Status} & \textbf{Feature}\\  \cline{1-7}
%\midrule
%\multirow[t]{4}{0.075\textwidth}{ } 
     &		&    Vostok 1 	& 1961  & USSR & retired & world's first \\  \cline{3-7}
 	& 	& Freedom 7 & 1961  & USA & retired & reached 187 km\\ \cline{3-7}
 	& 	& Apollo 11 & 1969  & USA & retired & \tabincell{c}{world's first crewed\\
 	moon landing; command  \\and service module \\ and lunar module }\\ \cline{3-7}
 	& spacecraft	& Shenzhou 5 & 2003  & China & retired & China's first\\ \cline{3-7}
 	&   & \tabincell{c}{Crew Dragon\\Resilience} & 2020 & \tabincell{c}{USA \\ (SpaceX)} & \tabincell{c}{ISS \\ transportation \\ mission\\ finished} & \tabincell{c}{world's first crewed\\spaceflight operated \\by a commercial entity;\\ reusable spacecraft} \\ \cline{3-7}
    &   & \tabincell{c}{Virgin Galactic\\Unity 22} & 2021 & \tabincell{c}{USA \\ (Virgin Galactic)} & \tabincell{c}{continued} & \tabincell{c}{air-launched suborbital \\spaceplane for \\tourism; Mach 3.2 by\\rocket engine; \\reached 86.1 km\\ }\\ \cline{3-7}
	&   & New Shepard 4 & 2021 & \tabincell{c}{USA \\ (Blue Origin)} & \tabincell{c}{continued} & \tabincell{c}{suborbital, reached\\ 107.05 km; first \\flight with owner;\\reusable spacecraft} \\ \cline{2-7}
 Crewed	& space shuttle	& Columbia & 1981  & USA & retired & \tabincell{c}{reusable \\ 23-ton payload} \\ \cline{2-7}
 	& 	& Salyut & 1971  & USSR & deorbited & world's first\\ \cline{3-7}
 	& 	& Skylab & 1973  & USA & deorbited & 360 m$^{3}$, 2249 days\\ \cline{3-7}
 	& 	& Mir & 1986  & USSR/Russia & deorbited & 350 m$^{3}$, 5511 days\\ \cline{3-7}
 	& space station	& \tabincell{c}{International\\Space Station \\(ISS)} & 1998  & \tabincell{c}{USA/Russia/ \\ESA$^{*}$/\\Canada/Japan} & orbiting & \tabincell{c}{915.6 m$^{3}$, \\ largest area; \\ longest service} \\ \cline{3-7}
 	& 	& Tiangong-1 & 2011  & China & deorbited & \tabincell{c}{China's first prototype }\\ \cline{3-7}
 	& 	& Tiangong & 2021  & China & orbiting & a space of 110 m$^{3}$ \\ \cline{1-7}
%\midrule
	& 	& Sputnik 1 & 1957 & USSR & deorbited &  \tabincell{c}{world's first\\artificial satellite} \\ \cline{3-7}
	& \tabincell{c}{Earth-orbit \\satellite}	&  \tabincell{c}{Hubble Space \\ Telescope (HST)} 	& 1990  & USA & orbiting & \tabincell{c}{numerous scientific \\findings including\\research leading to\\Nobel Prizes} \\ \cline{3-7}
& 	&  \tabincell{c}{Starlink} 	& 2019  & \tabincell{c}{USA \\ (SpaceX)} & orbiting & \tabincell{c}{largest satellite \\constellation for \\broadband access} \\ \cline{2-7}
& lunar probe	&  \tabincell{c}{Luna 1} 	& 1959  & \tabincell{c}{USSR} & finished & \tabincell{c}{world's first flyby} \\ \cline{2-7}
& solar probe	&  \tabincell{c}{Parker Solar \\Probe} 	& 2018  & \tabincell{c}{USA} & continued & \tabincell{c}{the first to enter\\ the solar atmosphere } \\ \cline{2-7}
& 	&  \tabincell{c}{Mariner 4} & 1964  & \tabincell{c}{USA} & finished & \tabincell{c}{world's first flyby} \\ \cline{3-7}
Uncrewed & Mars probe	&  \tabincell{c}{Perseverance} & 2020  & \tabincell{c}{USA} & continued & \tabincell{c}{Ingenuity, the first \\Mars robotic helicopter} \\ \cline{2-7}
& \tabincell{c}{}	&  \tabincell{c}{Voyager 1} & 1977  & \tabincell{c}{USA} & traveling & \tabincell{c}{first flybys of Jupiter, \\ Saturn and Titan; \\Furthest and first \\interstellar spacecraft } \\ \cline{3-7}
& \tabincell{c}{planetary \\ probe}	&  \tabincell{c}{Cassini–\\Huygens\\} & 1997  & \tabincell{c}{USA/ESA/\\Italy} & finished & \tabincell{c}{first Saturn orbiter\\and moons flybys;\\Huygens probe landed\\on Titan, first (non-\\Earth) moon landing\\  } \\ \cline{3-7}
& \tabincell{c}{}	&  \tabincell{c}{New Horizons} & 2006  & \tabincell{c}{USA} & continued & \tabincell{c}{first Pluto flyby } \\ \cline{2-7}
& \tabincell{c}{comet probe}	&  \tabincell{c}{Rosetta} & 2004  & \tabincell{c}{ESA} & deorbited & \tabincell{c}{first comet landing } \\ \cline{2-7}
& \tabincell{c}{planetary\\defense}	&  \tabincell{c}{Double Asteroid \\Redirection Test \\(DART)} & 2021  & \tabincell{c}{USA/ESA/\\Italy/Japan} & traveling & \tabincell{c}{first kind for testing a \\method against near-\\Earth objects (NEO)} \\ \cline{1-7}
%\bottomrule
\end{tabular}
   \begin{tablenotes}
        \footnotesize
        \item[*] the European Space Agency and it consists of 22 countries.
      \end{tablenotes}
\end{threeparttable}
\end{table*}

For the crewed spacecraft, the governmental bodies played a key role initially. Apollo 11 mission has made a critical historic accomplishment of landing mankind for the first time on another celestial body in 1969. The first and biggest space station that has been collaboratively built and operated by several countries, the International Space Station (ISS), has been in service since 1998 and has been visited by 244 persons from 19 nations. The lunar module of Apollo 11 mission and the ISS are illustrated in Figure~\ref{MoonISS} \footnote{This file is in the public domain in the United States because it was solely created by NASA. NASA copyright policy states that "NASA material is not protected by copyright unless noted". NASA copyright policy also applies to some other images used in this paper.}. However, a noticeable trend in recent years is that the commercial space industry fueled by private capital has exponentially grown to accelerate the technological transformation. One of the most influential corporations is Space Exploration Technologies Corp. (SpaceX) headquartered in Hawthorne, California. 

\begin{figure}%[H]
\centering
\subfigure[]{\includegraphics[width=8.2 cm]{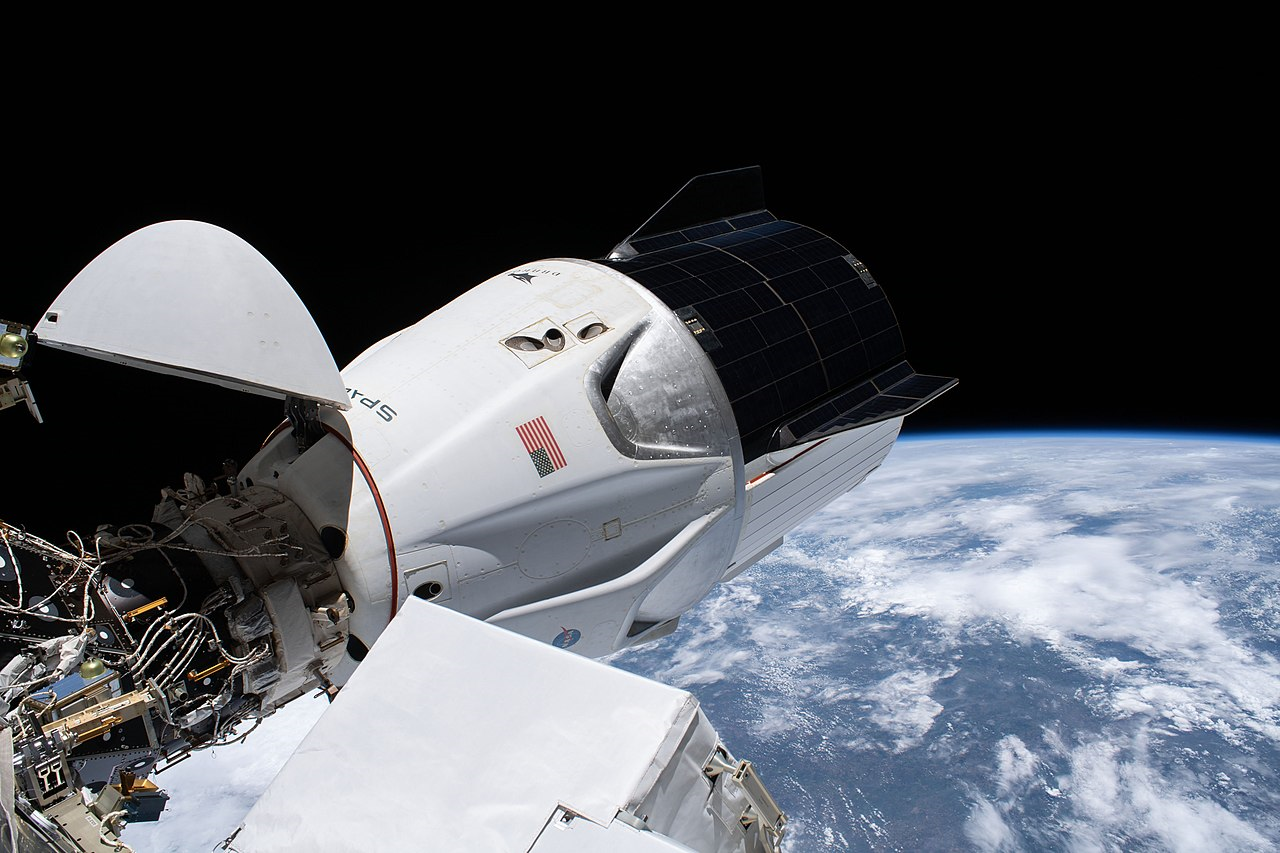}}
\subfigure[]{\includegraphics[width=5 cm]{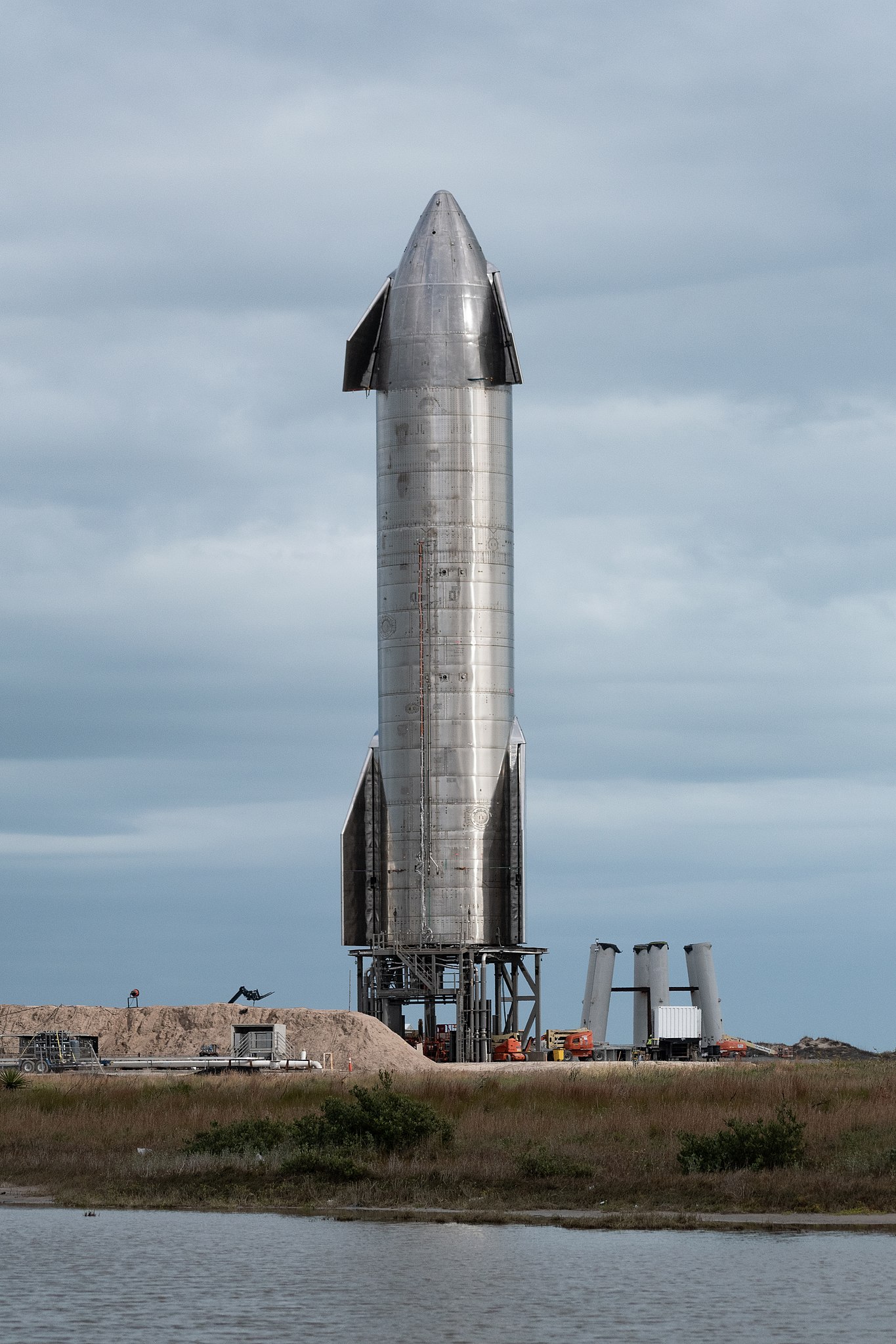}}
\caption{(\textbf{a}) The SpaceX Crew Dragon spacecraft docked to the ISS in Jan. 2021 \cite{ref-SpaceXCrew}. (\textbf{b}) Starship Serial Number (SN) 9 sitting on its launch pad in Starbase \cite{ref-Starship_SN9}.\label{SpaceX}}
\end{figure}

In 2015, SpaceX successfully launched and relanded a Falcon 9 rocket \cite{ref-webpage1} and launched a returned Falcon 9 later in 2017, which paved a promising road to low-cost reusable launching vehicles. Moreover, SpaceX's Falcon 9 booster that launched the Crew-1 mission with the Crew Dragon Resilience in November 2020 is being reused for the Crew-2 mission, which is the first time the same rocket booster has been used for multiple human launches. The SpaceX Crew Dragon and Starship spacecraft are illustrated in Figure~\ref{SpaceX}. Other non-governmental aerospace companies include California-based Virgin Galactic, which develops commercial spacecraft and provides suborbital spaceflights to space tourists, and Kent (Washington) headquartered Blue Origin, which also develops reusable launching vehicles and orbital technology.   

Furthermore, more uncrewed spacecraft have been launched, including various types such as Earth-orbit satellites, lunar/solar probes, planetary probes, comet probes, and asteroid probes (for planetary defense). The Hubble Space Telescope (HST) is a space telescope launched into the LEO in 1990 and is still in operation. It features a 2.4-m  mirror and is located at a variable altitude of around 537 to 540.9 km, with five main instruments to observe in the ultraviolet (UV), visible, and near-infrared regions of the electromagnetic spectrum. Many Hubble observations have led to breakthroughs in astrophysics \cite{ref-dark}, \cite{ref-how}, e.g., it assisted Scientists to determine the expansion rate of the universe and resulted in the Noble Prizes in 2011. Its end of mission is estimated to be around 2030--2040, while its successor, James Webb Space Telescope (JWST) was launched on Dec. 25, 2021, and is supposed to be in the orbit of Sun-Earth $\text{L}_{\text{2}}$ \cite{ref-Is}, \cite{ref-Space}, the second Lagrange point.   

\begin{figure} %[H]
\centering
\subfigure[]{\includegraphics[width=8.5 cm]{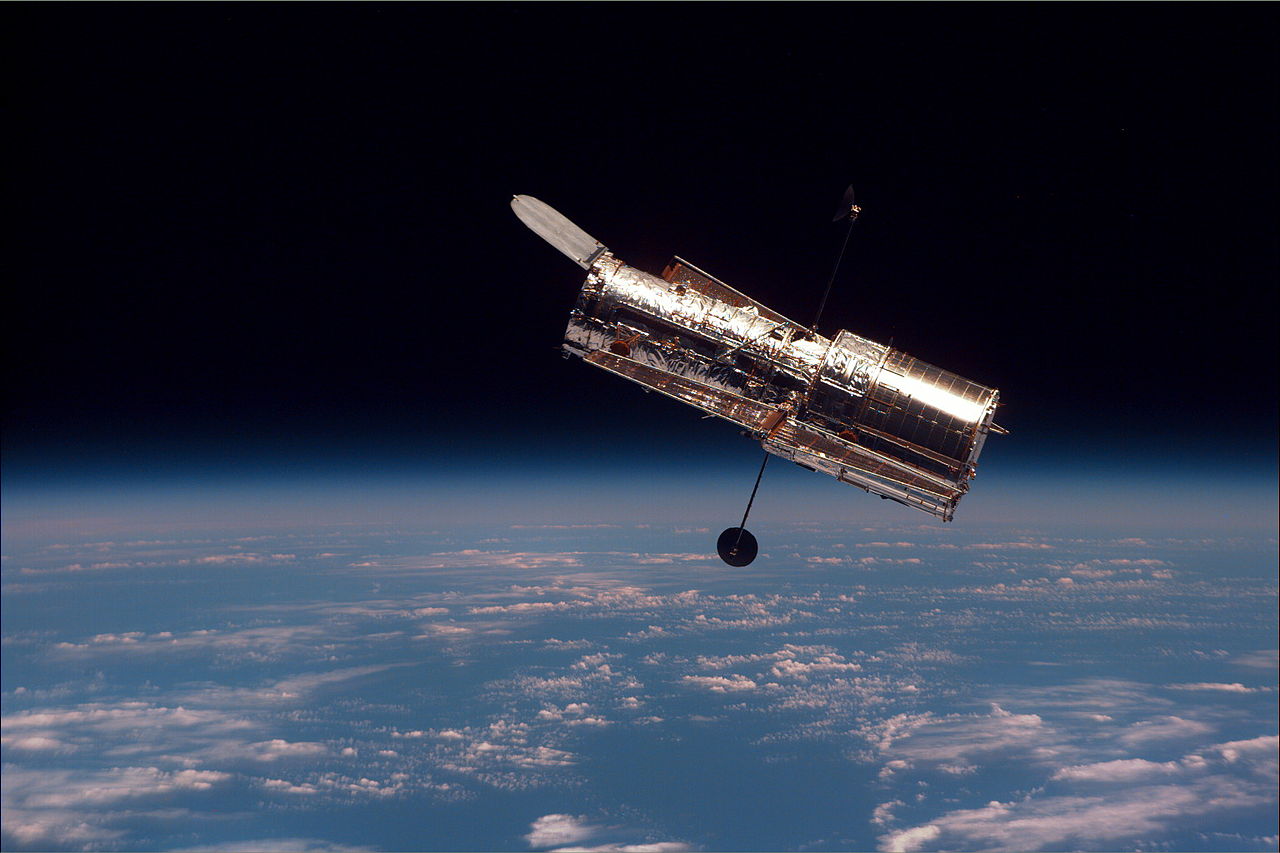}}
\subfigure[]{\includegraphics[width=8.5 cm]{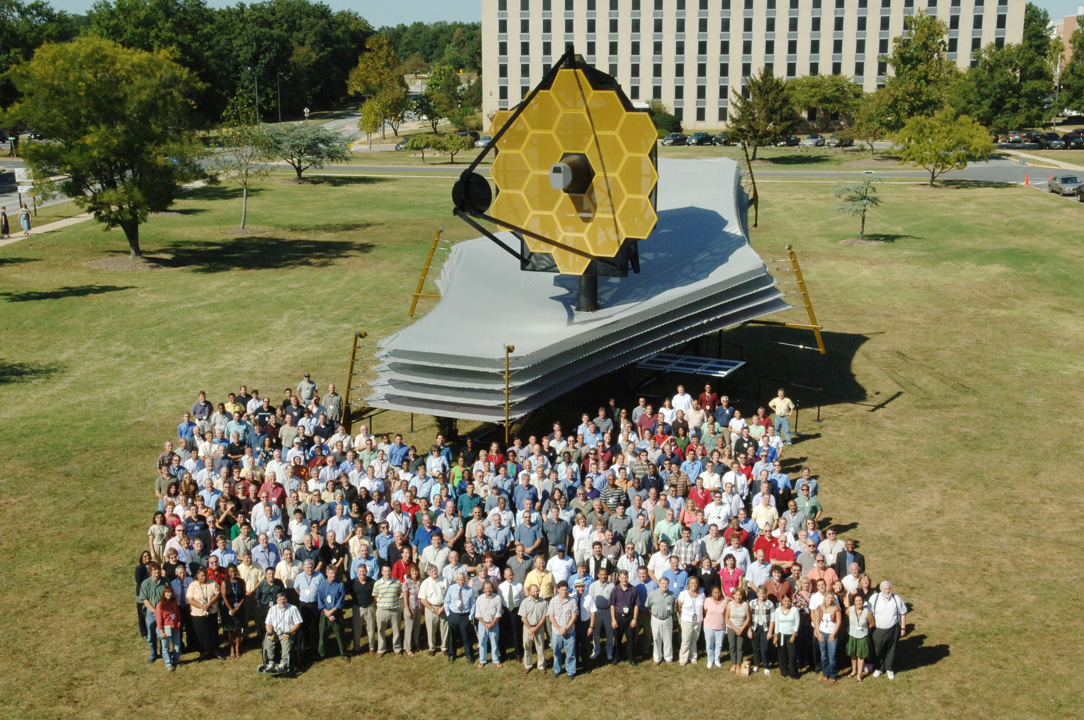}}
\caption{(\textbf{a}) Hubble as seen from the Space Shuttle \emph{Discovery} during its second servicing mission \cite{ref-Hubble}. (\textbf{b}) The James Webb Space Telescope full-scale model assembled on the lawn at Goddard Space Flight Center with the JWST team \cite{ref-JWST}.\label{Hubble-JWST}}
\end{figure}  

On the other hand, the planetary probes have been under the spotlight these years, and many nations have joined the race of sending probes to other planets. The recent operational Mars probes include Perseverance from the USA in 2020, Tianwen-1 from China, and Hope from United Arab Emirates (UAE). One of the earliest pioneering spacecraft of this type is Voyager 1, launched in Sep. 1977. As of Nov. 2021, it has already traveled 155.5 astronomical units (AU) (23.26 billion km), which is currently out of the heliosphere and has become the first interstellar spacecraft and most distant artificial object from Earth.

\begin{figure} %[H]
\centering
\subfigure[]{\includegraphics[width=8.18 cm]{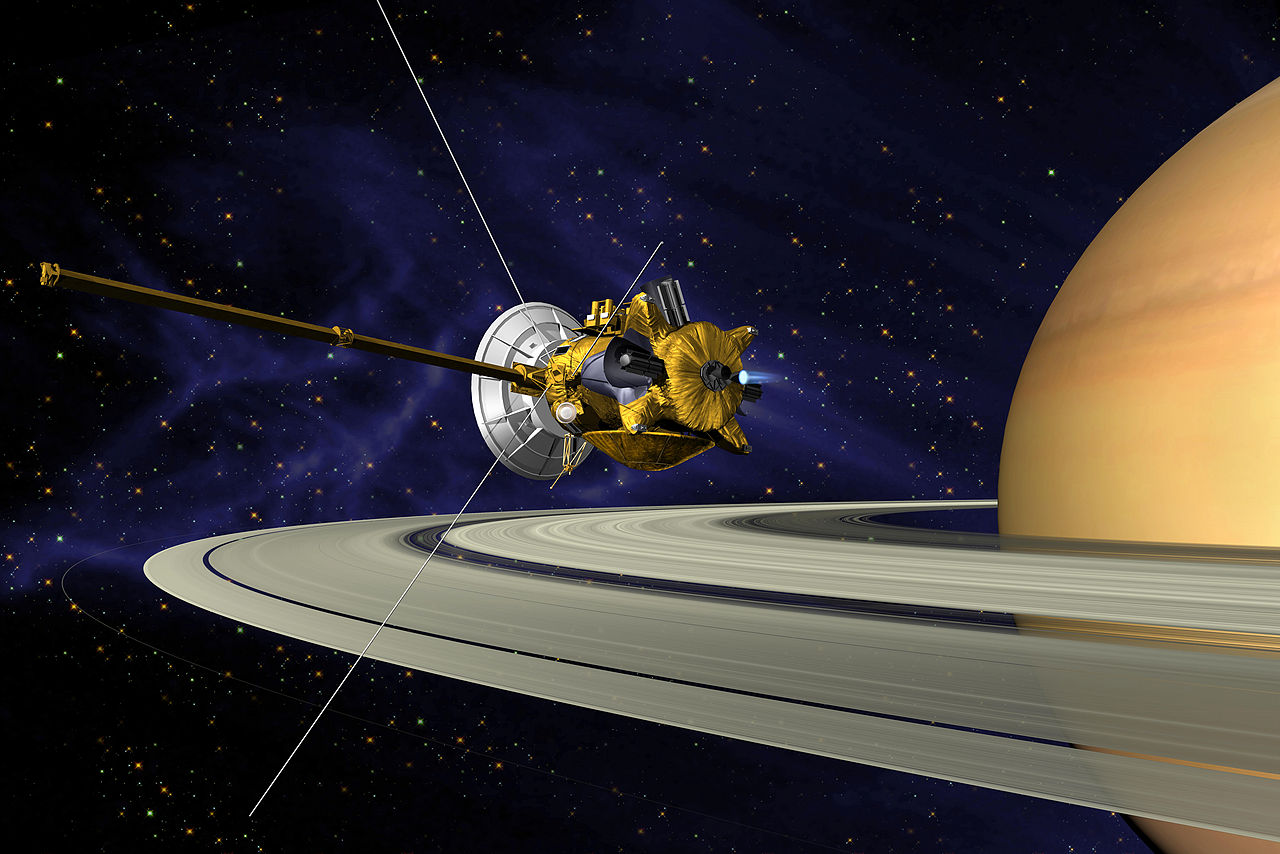}}
\subfigure[]{\includegraphics[width=6.18 cm]{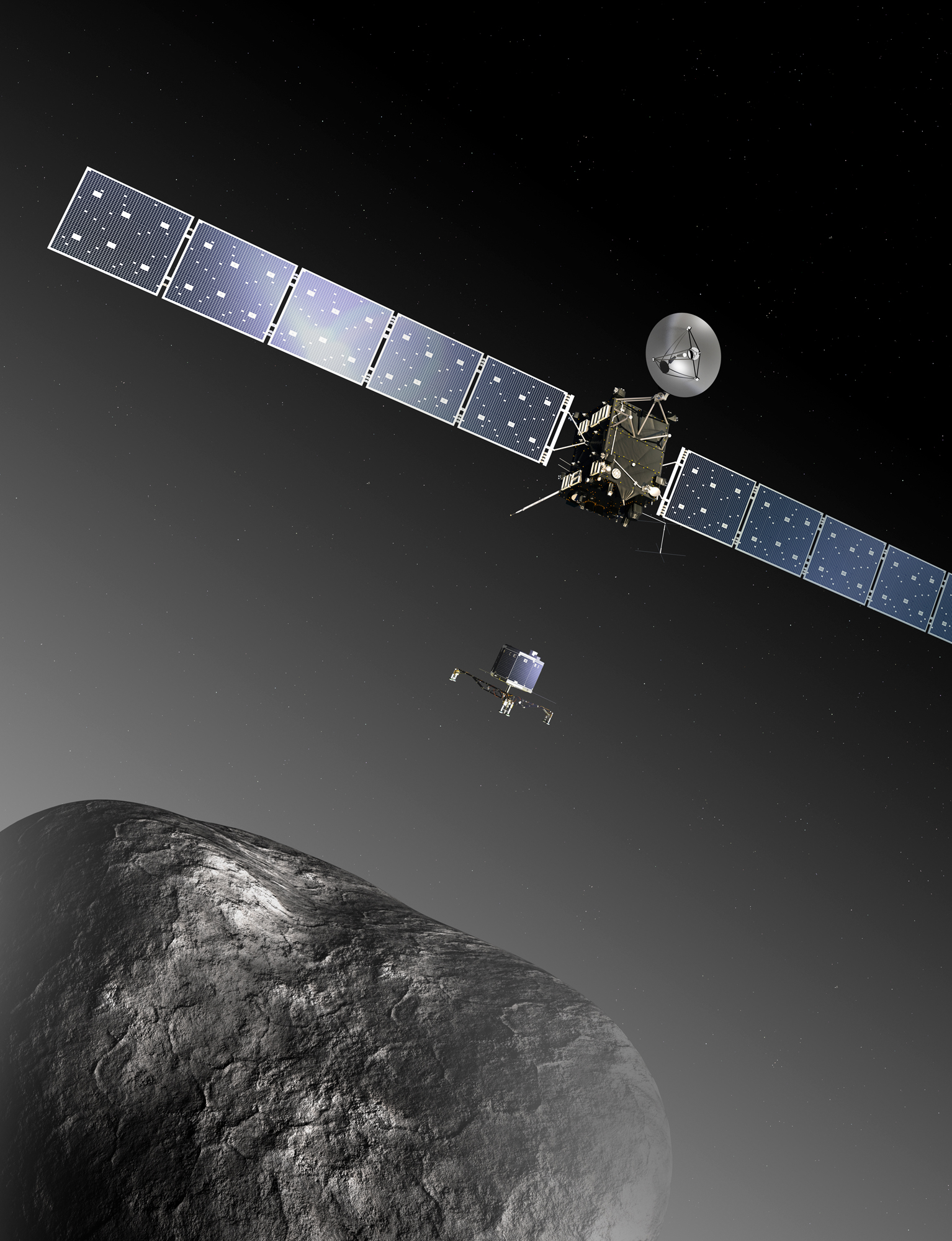}}
\caption{(\textbf{a}) Illustration of an artist's concept of Cassini-Huygens' orbit insertion around Saturn \cite{ref-Cassini}. (\textbf{b}) Artist’s illustration of the Rosetta orbiter deploying the Philae lander to comet 67P/Churyumov–Gerasimenk \cite{ref-Rosetta}.\label{Cassini-Rosetta}}
\end{figure}

The Cassini–Huygens spacecraft is a collaboration among NASA, the European Space Agency (ESA), and the Italian Space Agency (ASI) to study the planet Saturn and its system, including its rings and natural satellites. It consists of the Cassini space probe and ESA's Huygens lander, which landed on Saturn's largest moon, Titan, in Jan. 2005. An illustration is presented in Figure~\ref{Cassini-Rosetta} (a). NASA had conducted the mission extension after 2008 until Sep. 2017 when Cassini was intentionally sent into Saturn's atmosphere to be destroyed in order to prevent biological contamination. Many astonishing scientific discoveries have been reported from its 20 years journey in the space, which include the observation and analysis of Enceladus' water vapor plume \cite{ref-Enceladus}, \cite{ref-Water} and tests of Albert Einstein's general theory of relativity \cite{ref-Einstein}.

Except for the planetary probe, the comet probe is another type, and one representative is the Rosetta spacecraft designed and launched in Mar. 2004 by ESA to study comet 67P \cite{ref-Spiralling}. It was the first spacecraft to orbit a comet nucleus and the first spacecraft to be in close proximity of a frozen comet towards a direction warmed by the Sun. Rosetta orbiter dispatched the Philae lander for the first controlled touchdown on a comet nucleus. An illustration is shown in Figure~\ref{Cassini-Rosetta} (b). The Philae's instruments obtained the first images from a comet's surface and made the first on-site analysis of its composition. On Sep. 30, 2016, Rosetta was guided down to the comet's surface and the mission ended on impact \cite{ref-Historic}, \cite{ref-Mission}.  

Moreover, on Nov. 24, 2021, SpaceX Falcon 9 successfully launched the Double Asteroid Redirection Test (DART) spacecraft to test a method of planetary defense against near-Earth objects (NEOs) \cite{ref-Double}. The DART spacecraft hosts no scientific payload but sensors and a navigation system and plans to arrive at 65803 Didymos, a sub-kilometer asteroid and synchronous binary system classified as a potentially hazardous asteroid. Then, DART is set to crash into the asteroid system deliberately. The collision is scheduled in Sep./Oct. 2022 when the impact of 500 kg DART will target the center of Dimorphos, which is the minor-planet moon of 65803 Didymos. The ASI's secondary spacecraft, called LICIACube (Light Italian CubeSat for Imaging of Asteroids), a small CubeSat, will ride on DART and then separate ten days before impact to acquire images of the impact and ejecta as it drifts past the asteroid. The effects of the impact by DART will also be monitored from ground-based telescopes and radar.

In addition, a mission to 16-Psyche, which is a massive M-type asteroid with a mean diameter of 220 $\pm3$ km \cite{ref-VLT} and orbits the Sun in the main asteroid belt, was proposed to NASA in 2014. The concept of robotic Psyche orbiter was proposed by a team led by Lindy Elkins-Tanton \cite{ref-Elkins}, at Arizona State University. They proposed to use the spacecraft to orbit Phyche for 20 months and study its topography, surface features, gravity, magnetism, and other characteristics since Pysche is the only metallic core-like body discovered so far. The mission's launch date was moved up to July 2022 from the original date in October 2023, targeting a more efficient trajectory with a Mars gravity assist in 2023 and arriving in 2026. On Feb. 28 2020, SpaceX won a USD 117 million contract from NASA to launch the Psyche spacecraft in July 2022.

Apart from the representative as mentioned above missions, either accomplished or ongoing, there are many other spacecraft and space missions under construction or planned. Generally, exploring and discovering more unknowns of the universe is one of the most fundamental drives of the space projects, which stems from the curiosity rooted in the human species. Moreover, the technical advances in the related areas of the space industry reduce the cost and enable some previous theories to become practical and commercial plans such as space tourism and extraterrestrial colonization. Nevertheless, along with the exciting opportunities in the space age come unprecedented challenges, some of which may even pose potential extinction-level disasters to the human species. In this article, we will continue to investigate the critical carried-on space missions/projects focusing on space colonization, which will be followed by the review and analysis of the asteroid threat and planetary defense. To cope with the known challenges and difficulties, a novel framework and network based on the internet of spacecraft is presented and analyzed. The contribution of this paper can be unfolded in several aspects as follows:

\begin{enumerate}
\item A comprehensive review of spacecraft and space missions of critical milestones in the human history is introduced; 

\item A thorough review, investigation, and analysis of the multi-planetary colonization projects are presented. The potential asteroid threat and planetary defense are investigated and analyzed;

\item A novel framework named Solar Communication and Defense Networks (SCADN) consists of a large number of distributed spacecraft which is cable of enabling the internet of distributed deep-space sensing, communications, and defense. An in-depth analysis is conducted to analyze how SCADN can increase the success rate of early detecting the asteroids/comets/objects jeopardizing the safety of Earth and other colonized planets/moons. Moreover, the SCADN-based strategies of preventing the outer-space objects triggered extinction events are also presented and discussed. 

\end{enumerate}

The remainder of this paper is organized as follows, Section 2 reviews and analyzes current multi-planetary colonization projects and plans, followed by a detailed review and in-depth analysis of the NEO threat of the asteroid and the planetary defense in Section 3. Furthermore, Section 4 proposes the SCADN to cope with the potential asteroid threat and facilitate extraterrestrial colonization. Eventually, Section 5 concludes this paper.        

\section{Extra-Terrestrialization: Make Life Multi-planetary}

It is generally believed that life on Earth began in the water and had been aquatic for billions of years. About 530 million years ago, sea creatures likely related to arthropods first began to make forays onto land \cite{ref-Early}. Terrestrial invasion is one important milestone in the history of life \cite{ref-Terrestrial}, and the evolution of terrestrial vertebrates started around 385 million years ago. Authors in \cite{ref-Massive} have suggested from their data that a massive increase in visual range occurred prior to the subsequent evolution of fully terrestrial limbs as well as the emergence of elaborated action sequences through planning circuits in the nervous system. The eye size is almost tripled when comparing the tetrapods with digited limbs that evolved with early lobe-finned fish. From simulation results of examined animals viewing objects through water, the eye size increase provides a negligible increase in performance while the eye size increase enables a large performance increase of viewing through the air. Therefore, the advance in the visual sensory plays a crucial role in terrestriality.  

Today, on one hand, mankind's sensory abilities in both microscopic and macroscopic scales have been constantly enhanced by the technological advancements, and on the other hand potential resources in outer space and other celestial bodies have driven humans to evolve into a spacefaring civilization \cite{ref-Making}, \cite{ref-Making2}, which can also increase the probability of mankind's survival and prosperity in the universe. We perhaps can call this transformation as 'Extra-Terrestrialization' in the history of life. 

\subsection{Space Transportation}
It's not long since humankind learned how to combat gravity, and the transportation between the Earth's ground and outer space is still expensive and challenging. The space industry has been conventionally a capital-intensive risky business \cite{ref-Business}. The launch cost of earliest launch vehicles such as the Vanguard rocket is~\$894,700 US dollars (USD) per kilogram when using the consumer price index (CPI) calculator to correct the cost to the reported date (in 2018) \cite{ref-Recent}. A figure below is made to summarize and illustrate more historical data \cite{ref-Recent}, \cite{ref-Transcost} of the launch cost dating back to 1957, with the CPI calculator used to correct the original data (in 2018) to the end of 2021. 

\begin{figure*}
\centering
\centerline{\includegraphics[width=30pc]{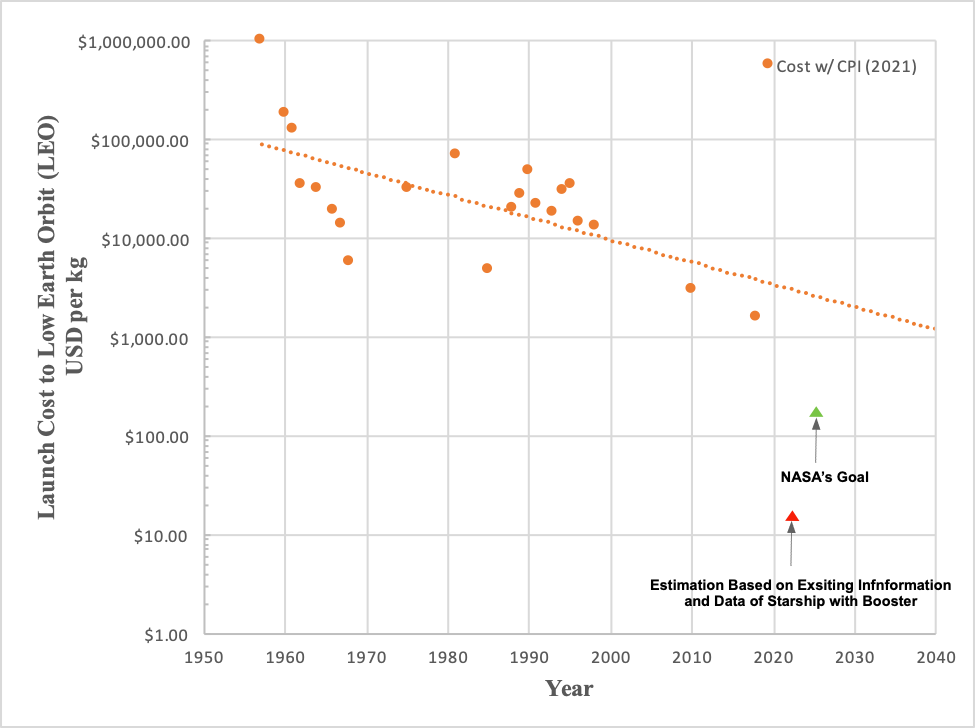}}
\caption{Low earth orbit launch cost change over time with 2021 CPI weighted.}\label{fig:launch}
\end{figure*}

As shown in  Figure~\ref{fig:launch},  the orange round dots indicate the launch costs to the LEO in the unit of USD per kilogram (USD/kg). The maximum and minimum costs are separated by almost 1,000 times due to technological advancements and productivity improvement. Another critical factor accelerating this disruption is the aforementioned private capital invested in the space industry in recent years. Since its first commercial mission in 2013, SpaceX has led the commercial launch service. One critical milestone was marked in 2015 when SpaceX successfully launched and relanded its Falcon 9 rocket on the ground pad \cite{ref-Falcon}, which paves the road for the launch vehicles' reusability and even lower launch cost.  

A trendy line based on existing available data indicates the general tendency of the cost change over a 60-year time period, which is a bit distant above the green triangular representing NASA's goal set to make the cost down to 100 USD/pound or 220 USD/kg in 2025 \cite{ref-NASA2025}. However, the scheduled first orbital flight of SpaceX's Starship that can deliver more than  100  tons (150 tons \cite{ref-Making})  of payload to the  LEO \cite{ref-Starship},  might bend the trendy line significantly more downwards in 2022. According to Elon Musk, the CEO and principal designer, the launch cost is estimated to be 2 million USD per launch \cite{ref-SpaceX2}, which can significantly reduce the cost to 20 USD/kg. 

However, the listed costs in the table are only for the launch reaching the LEO, while the trip to Moon and Mars will cost significantly more. Take Falcon 9, for example, the mass of payload to LEO is 22.8 t and will be reduced to 8.3 t and 4 t for reaching geosynchronous transfer orbit (GTO) and Mars transfer orbit (MTO), respectively. Correspondingly, the cost of GTO and MTO will be 2.75 and 5.7 times higher than LEO. The reusability, large capacity, and advanced massive fabrication of launch vehicles could significantly lower the cost.  

\begin{figure*}
\centering
\centerline{\includegraphics[width=40pc]{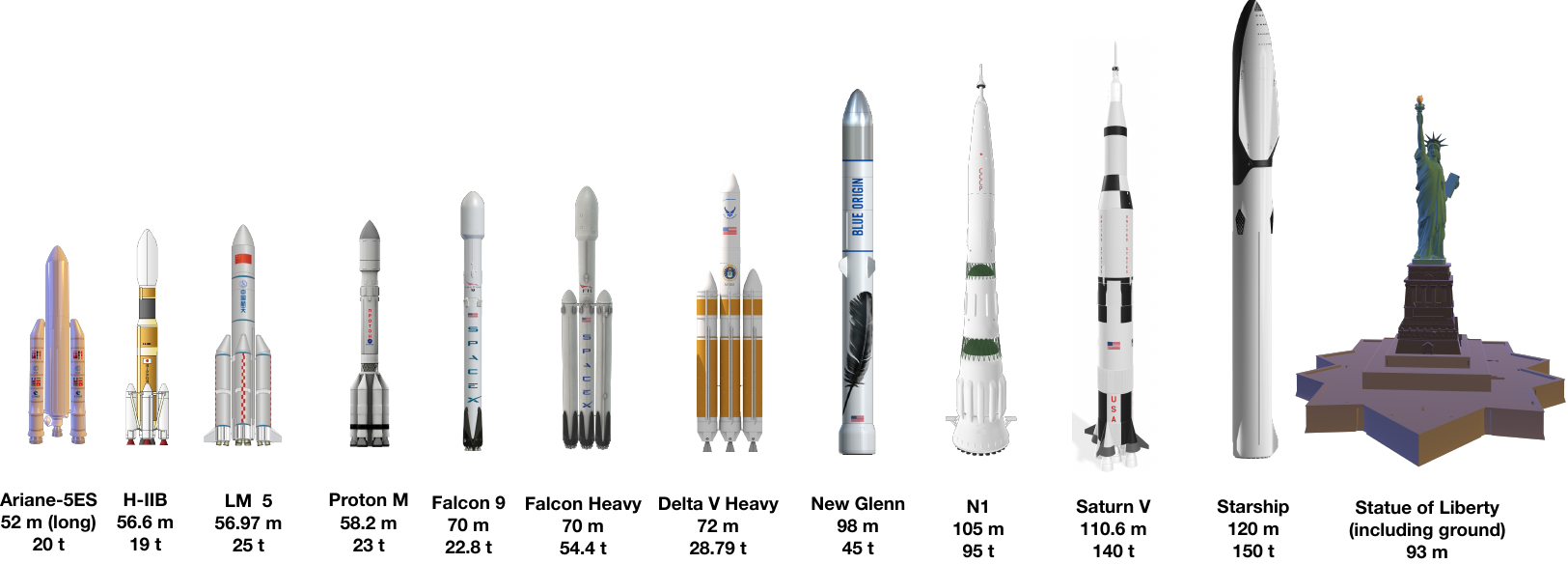}}
\caption{Payload to LEO comparison of several worldwide launch vehicles \cite{ref-Making2}.}\label{fig:WorldsRocket}
\end{figure*}

A payload comparison is summarized in \cite{ref-SpaceX2}, and it is worth mentioning that Big Falcon Rocket (BFR) was renamed after Nov. 2018 by SpaceX and the two-stage vehicle is technically composed of two parts, Starship (spacecraft) and the Super Heavy rocket (booster). As observed from the comparison in Figure~\ref{fig:WorldsRocket}, the Starship (with booster) is the most capable launch vehicle in human history and even outperforms SATURN V by >7\%. All well-known rockets are presented and compared in height and payload.  

\subsection{An Interplanetary Spacecraft Design Example: Starship}
Starship (without the booster) is 48-m long and its dry mass is around 85 tons. The ship will contain 1,100 tons of propellant with an ascent design of 150 tons and return mass of 50 tons \cite{ref-Making2}. SpaceX has designed the engine section in the rear, the propellant tanks in the middle and a eight-story tall payload bay in the front. In particular, the giant deep cryo liquid oxygen tank made of carbon fiber is 1,000 cubic meters of volume inside and more pressurized volume than an A380 airliner \cite{ref-Making2}. The delta wing at the back will enable Starship to handle the balancing challenges under various situations including zero/thin/dense atmosphere and/or with zero/low/heavy payload in the front \cite{ref-Making2}. Eventually, Starship can facilitate almost all types of entry/landing challenges in either a moon or a planet in the solar system.

Moreover, Starship (without booster) is expected to carry sub-cooled 240 tons of methane (CH) and 860 tons of oxygen as the propellant. The ship engine section consists of four vacuum Raptor engines and two sea-level engines, and all six engines are capable of gimbaling \cite{ref-Making2}. However, in another reference \cite{ref-Making2}, the interplanetary spaceship is specified to have 3 sea-level engines with 6 vaccum engines. There might be more starship design versions when the deployment and testing are being carried on. Moreover, the booster, or Super Heavy rocket is equipped with 42 Raptor engines. The main body of the whole Starship including Super Heavy rocket is currently made of SAE 304L stainless steel thanks to its low cost, high melting point, strength at cryogenic temperature, and ease of manufacturing. The heat shield attached to the ship body is composed of thousands of black hexagon tiles that contain silica. By using hexagon tiles, hot plasma on atmospheric entry cannot accelerate through zigsag gaps and the heat shield can withstand up to 1,400 °C \cite{ref-Hexagon}, and more importantly they can be used many times without maintenance.   

Furthermore, the Mars transit configuration (MTC) of Starship consists of 40 cabins that can hold more than 100 people per flight taking approximately 6 months (3 months in very good scenario \cite{ref-Making2}). In addition, Starship-MTC is expected to facilitate life-support and other requirements of a long interplanetary trip such as storage, entertainment, and solar storm shelter which can resist the solar flare which is the abrupt eruption of electromagnetic (EM) radiation in the Sun's atmosphere and the resulted energetic protons can pass through the human body, causing biochemical damage. 

\begin{figure*}%[H]
\centering
\centerline{\includegraphics[width=42pc]{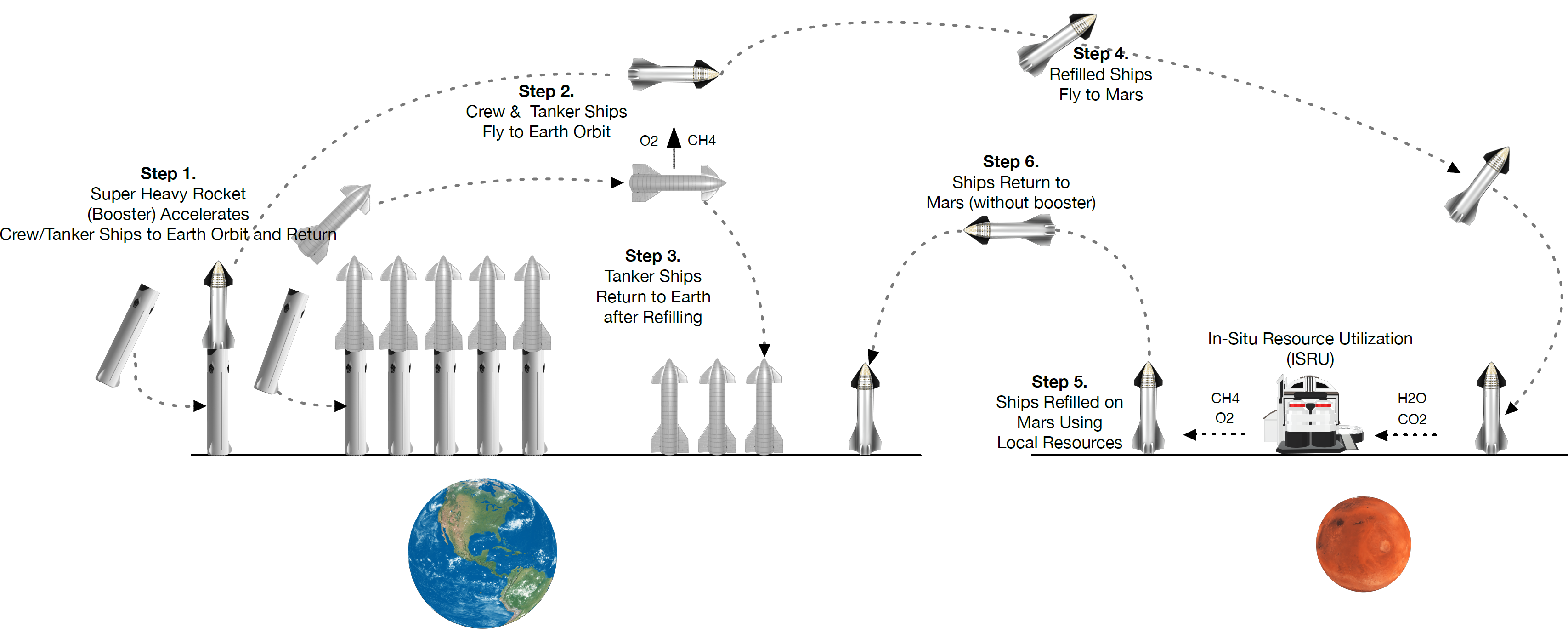}}
\caption{Illustration of Mars transit system architecture and flow chart, based on \cite{ref-Making}.}\label{fig:MarsTransport}
\end{figure*}%[H]mar

Take the Mars transportation, for example, the system architecture and flow chart are illustrated in Figure~\ref{fig:MarsTransport} based on \cite{ref-Making} \cite{ref-Making2}. The Super Heavy rocket launches the crewed and/or uncrewed Starship spacecraft to the Earth orbit first, then is refilled by uncrewed Starship tankers that are already deployed in the Earth orbit in advance \cite{ref-Making}, \cite{ref-Making2}. Refilling the propellant (liquid oxygen (O$_{2}$) and methane (CH$_{4}$)) in the orbit is essential as it can maximize the spaceship payload on the trip to the final destination such as Moon or Mars, which eventually could reduce the vehicle size and cost by 5-10 times and increases the launch rate \cite{ref-Making}. It is worth mentioning that establishing a self-sustaining base on Moon or Mars or other celestial bodies, a large number of spacecraft likely to be several thousand with tens of thousands of refilling operations, which translates to many launches every day \cite{ref-Making2}. Then, the refilled Starships will continue the travel to Mars while the tanker refill ships will return to Earth for refilling and the next launch. 

\subsection{Foundation on Another Celestial Body}
Founding a human base or even a self-sustaining city on another celestial body is challenging in light of the current technological level \cite{ref-Mars}. Building the first colonization for humankind on another celestial body without or with a thin atmosphere, with more severe damage from cosmic rays, solar radiation, or even asteroid impact, has a challenging kickoff. Specifically, constructing reliable shelters either on the ground or underground is a huge mission requiring significant workforces and resources that depend on transportation at the very beginning. Take Mars, for example, 1 million people is a threshold required to maintain a civilization, which translates to at least 10,000 trips and order of 1,000 ships \cite{ref-Making}. Another challenge is, the Earth-Mars rendezvous timing is roughly 26 months when the distance between two celestial bodies becomes periodically minimum, which means a huge number of preparation and launch will occur in a very narrow time window and the Mars fleet have to depart en masse \cite{ref-Making}.

Compared to any other celestial body in the solar system, Mars is the most habitable telluric planet due to the suitable distance from the Sun and similar day length, land mass, and a bit less gravity (about 38\% of Earth). The average Mars atmospheric pressure is about 600 Pa, or 0.5921\% of Earth's standard atmosphere (ATM). Moreover, the Mars atmosphere is mainly composed of 95.32\% CO$_{\text{2}}$, 2.7\% N$_{\text{2}}$, 1.6\% Ar, 0.13\% O$_{\text{2}}$, etc. Moreover, billions of years ago, when Mars was warmer, and the atmosphere was denser, channel-like valleys and riverbeds were highly likely to emerge on the surface with flowing water \cite{ref-Warming} \cite{ref-Water2} \cite{ref-Seasonal}. 

However, due to the low temperature and atmospheric pressure today, liquid water cannot exist on the Mars' surface, although some water vapour is held in the atmosphere and some other liquid water might be released occasionally by the volcanic activity and asteroid impact. Most of the water on Mars is locked in the polar caps, of which the larger one is the northern cap, Planum Boreum, with a dimension of 1,000 km across and 1.2 km thick. Furthermore, the northern cap is composed of 90\% of water ice and the carbon dioxide (CO$_{\text{2}}$) ice that has a seasonal change observed by the Hubble space telescope \cite{ref-JPL} while the southern cap (Planum Australe) has a thick base of water ice topped with a 8-m layer of (CO$_{\text{2}}$) ice. The southern cap is the only place where CO$_{\text{2}}$ ice persists on Mars' surface year-around.  

The rich content of CO$_{\text{2}}$ and water ice on Mars can facilitate local mass production of propellant based on the Sabatier reaction process as shown in Figure~\ref{fig:MarsTransport}. For the oxygen and methane production, the Sabatier catalytic reactor (SR) and co-production/electrolysis are shown as follows \cite{ref-CalTech} \cite{ref-Methanation}:

\begin{equation*}
     \ce{ CO2 + 4 H_{\text{2}} -> CH_{\text{4}} + 2 H_{\text{2}}O}~\text{(g)}  ~~-164~\text{kJ~mol}^{-1}~(\text{at~298 K}),
     \label{Sabatier}
\end{equation*}
where Ruthenium (Ru) is used as catalyst under the temperature of 200--300 $^\circ$C. 

\begin{equation*}
     \ce{ 2 H2O_{\text{(l)}} ->[\text{electrolysis}] 2 H_{\text{2}}_{\text{(g)}} + O_{\text{2}}_{\text{(g)}}}.  
     \label{Sabatier2}
\end{equation*}

\begin{figure*} %[H]
\centering
\subfigure[]{\includegraphics[width=12 cm]{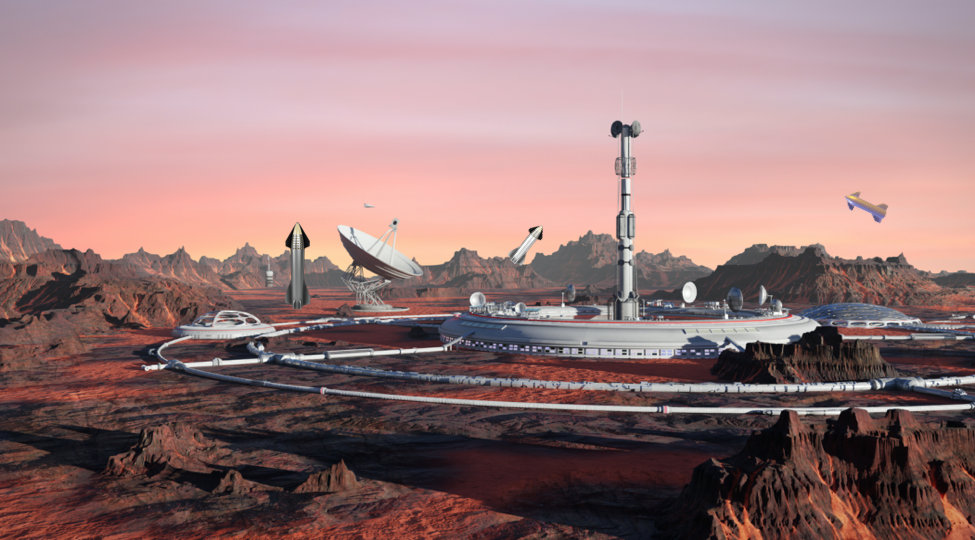}}\\
\subfigure[]{\includegraphics[width=12 cm]{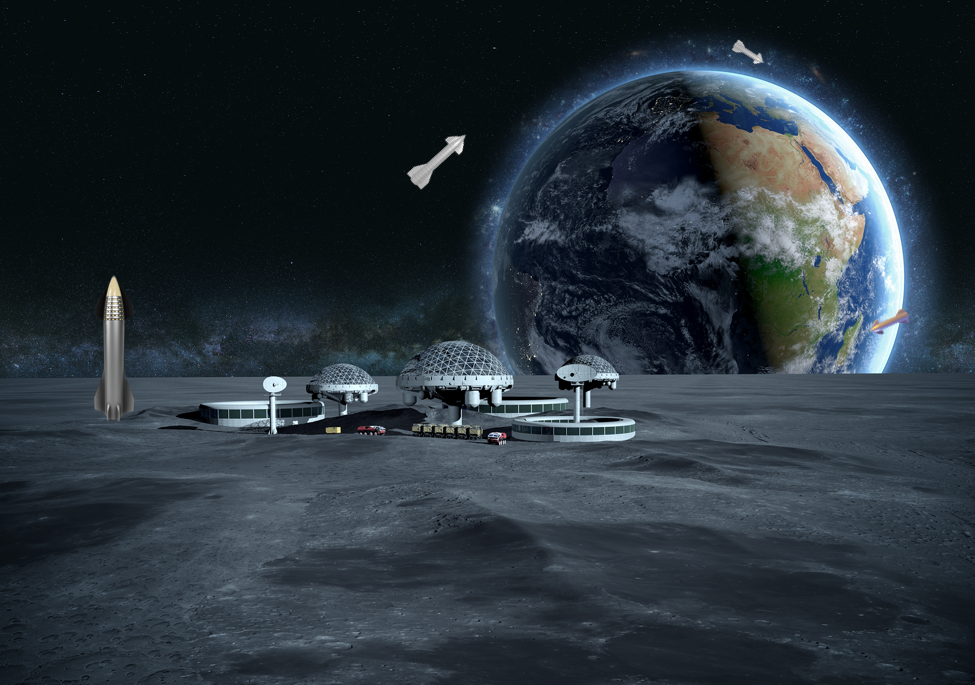}}
\caption{Illustration of foundation of (\textbf{a}) Mars city and (\textbf{b}) Moon base with massive connected spaceships.\label{MarsMoon}}
\end{figure*} 

According to \cite{ref-CalTech}, every 1 kg of propellant made on Moon or Mars saves 7.4 to 11.3 kg in the LEO. The Mars ISRU can significantly take advantage of atmospheric and soil resources and thus facilitates future Moon and Mars missions or even further missions to other celestial bodies. On the other hand, the minerals contained in the Mars soil such as Iron, Aluminium, Magnesium, Silicon, Titanium, etc. can be used for building self-sustaining cities and spacecraft. However, there are challenges such as how to operate the construction and mass production in extreme environments with extreme temperature, pressure, dust, cosmic rays radiation, low gravity and micro-gravity.

\begin{figure}%[H]
\centering
\centerline{\includegraphics[width=27pc]{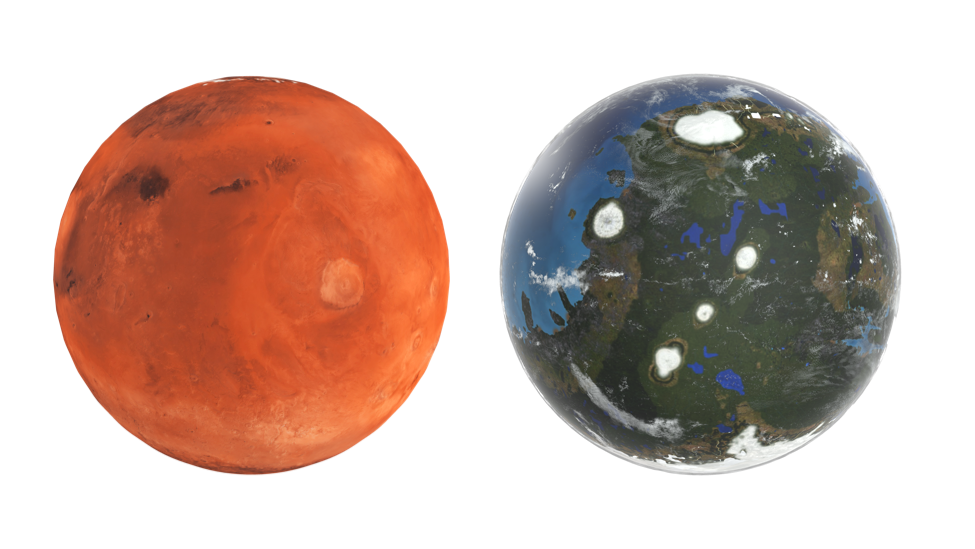}}
\caption{Illustration of Mars before and after terraforming.}\label{MarsTerraforming}
\end{figure}%[H]

Moreover, due to the major motivations of conducting scientific research, resource mining and ISRU \cite{ref-Moon}, the Earth's moon can be another crewed base although it is much smaller than a planet and got only 16.5\% gravity as Earth and no atmosphere. Moon got water ice, oxygen (in iron-rich lunar minerals and glasses) \cite{ref-Lunar}, and abundant solar power and mineral resources such as Iron, Aluminium, Magnesium, Silicon, Calcium, Titanium, and rare-earth elements that are used for manufacturing electric vehicles, wind turbines, electronic devices and clean energy technologies. More importantly, Moon is estimated to contain more than 1 million tons of helium-3 ($^{3}$He) to the surface \cite{ref-He3}, which can be potentially harvested for nuclear fusion. Furthermore, on a bigger picture, building the Moon base requires reliable systems that can remotely operate and accumulate considerable experiences for trips to Mars \cite{ref-MoonMars} and even further celestial bodies. As illustrated in Figure~\ref{MarsMoon}, a large number of (wirelessly) connected spaceships enabled with AI and advanced telecommunication and computing capabilities can mitigate the cross-planets/moons missions' challenges and enable a seamlessly cooperative transportation.

\subsection{Terraforming}
After setting the first foundation or even the first self-sustaining city with its own relatively comprehensive industry, agriculture and services. The terraforming, which is the hypothetical process of deliberately modifying the atmosphere, temperature, surface topography or ecology of a celestial body to make it more similar or habitable to the environment of Earth, should commence. It can be a very long procedure but will benefit, in the long term, mankind and even Earth's animals and plants for the multi-planetary transition, as depicted in Figure~\ref{MarsTerraforming}.  

In contrast to the controversy on terraforming Earth's Moon due to its higher difficulty \cite{ref-NeverMoon} \cite{ref-MoonTerraform}, Mars is the most likely candidate to be terraformed. For example, it is believed that Mars used to be a warmer and wetter planet \cite{ref-Water2} \cite{ref-Seasonal}, with a thicker atmosphere but lost it during a course of hundreds of millions of years due to reasons still unclear \cite{ref-Kargel}. Some likely mechanisms behind it could be related to the planetary surface absorption of greenhouse gases such as carbon dioxide, lack of powerful magnetosphere around Mars due to the ceased dynamo function \cite{ref-Mantle}, or asteroid impacts that ejected the ancient Martian atmosphere into the deep space \cite{ref-MarsStory}. However, there is evidence from the NASA MAVEN mission that the Mars atmosphere got removed mainly due to coronal mass ejection events but Mars does still retain a level of magnetosphere covering approximately 40\% of the surface rather than uniformly covering and protecting the atmosphere from solar wind \cite{ref-SolarWind}.   

Theoretically, terraforming Mars is the reverse procedure of how it lost the atmosphere and water. Among many proposals, including nuking Mars \cite{ref-Nuke}, several ones are particularly worth mentioning. The greenhouse gases such as carbon dioxide can be generated from Mars through heating \cite{ref-Faure}, to form CO$_{\text{2}}$ clouds to scatter the infrared radiation and thus trap the incoming solar radiation \cite{ref-Early2}. Subsequently, the raised temperature can add more greenhouse gases to the atmosphere, and the two processes will augment each other and form a positive loop. Moreover, authors in \cite{ref-Aerogel} have demonstrated a 2–3 cm-thick layer of silica aerogel which will simultaneously transmit sufficient visible light for photosynthesis, block hazardous ultraviolet radiation and raise temperatures underneath it permanently to above the melting point of water, without the need for any internal heat source. All experiments and modeling are under Martian environmental conditions. On the other hand, 
as motivated by the microorganisms' diverse roles in sustaining life on Earth, authors in \cite{ref-Microbes} have proposed a framework for new discussion based on the scientific implications of future terraforming such as avoiding accidentally distributing Earth's harmful microorganisms and genes to extraterrestrial areas and letting life stretch as a continuum connecting certain planetary bodies, with microbes doing the work of pioneer habitat conditioning. However, it requires many rigorous systematic, and controlled experimental studies with an ethical platform developed on Earth in advance.

Several other celestial bodies, such as Venus, and Mercury, have also been studied for terraforming. Still, Mars may be the most feasible one within mankind's technological capabilities, although the economic resources required can call for an enormous global joint effort in various aspects. 

\section{Space Threats and Planetary Defense}

\subsection{Serious Threats from Outer Space}

\subsubsection{Cretaceous–Paleogene Extinction}
Threats such as asteroids and comets from somewhere outside Earth can result in enormously catastrophic consequences. The Cretaceous–Paleogene (K–Pg) extinction event is also known as the Cretaceous–Tertiary (K–T) extinction and it was a sudden mass extinction of around 75\% of the plant and animal species on Earth, which happened approximately 66 million years ago \cite{ref-Time} \cite{ref-KTBoundary}. With the exception of some ectothermic species, no tetrapods weighing more than 25 kilograms survived. A large number of species were erased from Earth in the K–Pg extinction, and the best-known are the non-avian dinosaurs. Moreover, it has destroyed numerous other terrestrial organisms such as mammals, birds \cite{ref-Birds}, lizards and snakes \cite{ref-Lizards}, insects, plants, teleost fishes \cite{ref-Fishes}, pterosaurs, plesiosaurs and mosasaurs \cite{ref-KTBoundary}. 

The root cause of the K–Pg extinction event was originally proposed to be the impact of a massive comet or asteroid 10 to 15 km wide about 66 million years ago, by a team of scientists led by Nobel Prize-winning physicist Luis Alvarez and his son Walter. The impact hypothesis is also known as the Alvarez hypothesis and got supported by the discovery of the 180 km sized Chicxulub crater in the Gulf of Mexico's Yucatán Peninsula which was summarized in the publication of 1991 \cite{ref-Crater}. Moreover, authors in \cite{ref-Chicxulub} have presented more conclusive evidence that the K–Pg boundary clay represented the debris (unusually high levels of the metal iridium) from an asteroid impact. Such a massive impact would have instantly led to devastating shock waves, a large heat pulse, and global tsunamis. Moreover, the release of enormous quantities of dust, debris, and gases would have resulted in a prolonged cooling of Earth's surface, low light levels, and ocean acidification that would have decimated primary producers including microscopic marine algae, as well as those species reliant upon them \cite{ref-Chicxulub}.

In particular, there is broad consensus that the Chixculub impactor was an asteroid rather than a comet with a broadly accepted diameter of around 10 kilometers. The impact released estimated energy of between 1.3×$10^{24}$ and 5.8×$10^{25}$ joules (1.3–58 yottajoules), or 21–921 billion Hiroshima A-bombs \cite{ref-Durand}. Authors with expertise in modeling nuclear winter have published their research in 2013 about the impact winter \cite{ref-Reevaluation}, which indicated that the entire terrestrial biosphere might have burned based on the amount of soot in the global debris layer. Supposedly, a global soot-cloud blocked the sun and created a long-term winter effect. More recently, in 2020, authors have published the climate-modeling of the extinction event supporting the asteroid impact and instead of volcanism \cite{ref-Chiarenza}. However, the extinction also provided other opportunities to enable many groups to radiate. For example, mammals diversified in the Paleogene \cite{ref-Radiation} and evolved new forms such as horses, whales, bats, and primates which made the emergency of the Homo sapiens possible. 

\subsubsection{Other Significant Events}

Besides the serious K-Pg extinction event, some other significant events in the more recent timeline are also believed to be related to the comet or asteroid. The Younger Dryas that happened around 12,900 to 11,700 years before the present (BP) \cite{ref-Greenland} was a return to glacial conditions after the Late Glacial Interstadial (LGI), which temporarily reversed the gradual climatic warming after the Last Glacial Maximum (LGM) started fading around 20,000 BP. For example, within decades, a sudden decline of temperatures in Greenland by 4 to 10 °C took place \cite{ref-Greenland}. It is hypothesized that an impact event occurred in North America around 12,900 BP and initiated the Younger Dryas (YD) cooling \cite{ref-Megafaunal} that lasted for 1,200 years and became one of the possible major causes of the Quaternary extinction. Nanodiamonds which are usually the high-temperature products during the extraterrestrial collision \cite{ref-Temperature} have been found at the Younger Dryas Boundary (YDB) in the northern hemisphere \cite{ref-Nanodiamonds} \cite{ref-Kinze}. The catastrophic consequences of this impact including the abrupt YD cooling contributed to the late Pleistocene megafaunal extinction (e.g. woolly mammoth, saber-toothed predator), promoted human cultural changes, and resulted in immediate decline in some post-Clovis human populations \cite{ref-Megafaunal}.

When the timeline moves closer to the present, on the morning of June 30, 1908, an enormous explosion of around 12 megaton (of TNT) \cite{ref-Jenniskens} (approximately 800 Hiroshima A-bombs) occurred near the Podkamennaya Tunguska River in now Krasnoyarsk Krai, Russia. An estimated 80 million trees over a forest area of 2,150 km$^{2}$ were flattened \cite{ref-Farinella} with three people possibly dead in the event reported by witnesses \cite{ref-Jenniskens}. The atmospheric explosion of a stony meteoroid with an estimated dimension of 50–60 meters \cite{ref-DePater} happened at an altitude of around 5 to 10 kilometers after entering Earth's atmosphere with a high speed of about 27 km/s \cite{ref-Jenniskens} \cite{ref-Nemiroff}. The meteoroid is believed to have disintegrated and exploded at a similar altitude. Therefore there is no impact crater found. The sounds were accompanied by a shock wave that knocked people off and broke windows hundreds of kilometers away \cite{ref-Jenniskens}.

The explosion was registered at seismic stations across Eurasia, and air waves  were detected in Germany, Denmark, Croatia, the United Kingdom, and Washington, D.C. \cite{ref-Whipple}. Over the next few days, night skies in Asia and Europe were aglow \cite{ref-Watson}. Moreover, a Smithsonian Astrophysical Observatory program at the Mount Wilson Observatory in California observed a months-long decrease in atmospheric transparency and an increase in suspended dust particles \cite{ref-Turco}. The Tunguska event is the largest impact event on Earth in recorded history.

Within the last decade, the most significant meteor event happened on February 15, 2013 at about 03:20 UTC, over the southern Urals region of Russia. An around 20m-sized asteroid entered the atmosphere at a shallow 18.3 ± 0.4 degree angle with a speed relative to Earth of 19.16 ± 0.15 kilometers per second \cite{ref-Popova}. According to the record, the meteor generated brighter light than the Sun and was visible as far as 100 km, and some eyewitnesses also felt intense heat from the fireball. Its peak radiation occurred at 29.7 ± 0.7-km, then the fragmentation left a thermally emitting debris cloud in this period, the final burst of which occurred at 27.0-km altitude over Chelyabinsk Oblast. The Chelyabinsk meteor shone 30 times brighter than the Sun, and had an energy equivalent to more than 500 kilotons of TNT \cite{ref-Schiermeier} \cite{ref-500} which is around 33.3 times as the energy released from the Hiroshima A-bomb.

Furthermore, the object did not release all of its kinetic energy in the form of a blast wave as some 90 kilotons of TNT  of the total energy was emitted as visible light, according to NASA's Jet Propulsion Laboratory (JPL) \cite{ref-Yeomans}. The explosion has resulted in injuring about 1,500 people (a large majority due to shattered glass) and the damage to 7,200 buildings in six cities across the region. It is the largest known natural object to have entered Earth's atmosphere since the 1908 Tunguska event, and the only meteor confirmed to have resulted in many injuries.

\begin{figure*} %[H]
\centering
\centerline{\includegraphics[width=32 pc]{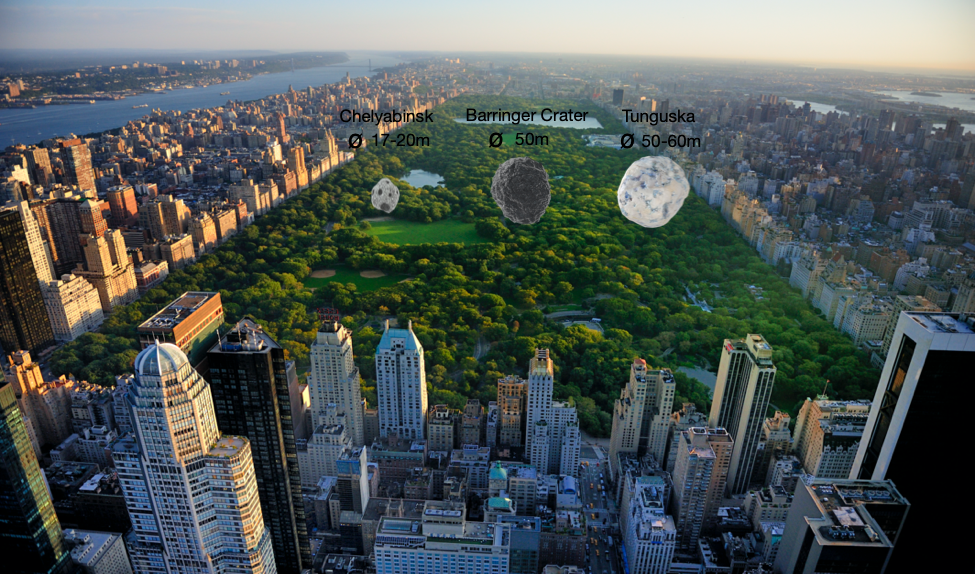}}
\caption{Comparisons of approximate sizes of three notable impactors with New York Manhattan as the background.}\label{Asteroid}
\end{figure*}

In terms of authors in \cite{ref-Borovicka}, they have particularly pointed out that Chelyabinsk meteor's orbit was incredibly similar to the 2.2-km-diameter near-Earth asteroid 86039, which was first observed in 1999. The unusual similarity strongly suggests that the two bodies may be used to be part of the same object. According to the simulations, the asteroid spent six weeks before the impact within an elongation of 45-degree from the Sun, which is a region of the sky, giving no access to ground-based telescopes. At earlier times, the asteroid was always too faint to be observed \cite{ref-Borovicka}. As further illustrated in Figure~\ref{Asteroid}, the sizes of Chelyabink meteor, Tunguska meteor, and Barringer crater meteor \cite{ref-Melosh} are demonstrated and compared.  

%[H]

% http://theory.caltech.edu/~kapustin/karandash/w5.pdf

%https://science.nasa. gov/science-news/science-at-nasa/2008/30jun_tunguska#:~:text=At%207%3A17%20a.m.%20(local,impact%20crater%2C%22%20said%20Yeomans.

% https://en.wikipedia.org/wiki/Tunguska_event
%https://en.wikipedia.org/wiki/Asteroid_impact_avoidance

\subsection{Planetary Defense}

\subsubsection{Timeline}

As early as in the book published in 1964 \cite{ref-Cole}, authors Dandridge M. Cole and Donald W. Cox mentioned the potential risks and dangers of planetoid impacts, both naturally and those that might be intentionally brought about with hostility. Moreover, they argued for categorizing the minor planets and developing the technologies to land on, deflect, or even capture planetoids. In 1967, students at MIT did a design study called Project Icarus to prevent a hypothetical impact on Earth by asteroid 1566 Icarus \cite{ref-Day} with considerable publicity received. Furthermore, a series of studies on the historical impact events on Earth such as \cite{ref-Alvarez} \cite{ref-Turco} in the 1980s lead to a later program to map objects in the Solar System which cross Earth's orbit and are large enough to cause serious damage if they impact on Earth.

In Jan. 1992, NASA sponsored a near-Earth-object Interception Workshop hosted by Los Alamos National Laboratory when the issues and challenges to cope with intercepting celestial bodies that could impact Earth were discussed \cite{ref-Canavan}. Furthermore, a 1992 US Congressional study produced a Spaceguard Survey Report, and it resulted in a 1994 mandate that NASA locates 90\% of near-Earth asteroids larger than 1 km within 10 years. The impact of a celestial body much larger than 1 km diameter could lead to worldwide damage potentially including the extinction of the human species. The initial Spaceguard goal was achieved with a period slightly longer than 10 years. An extension to the project required NASA to reduce the minimum diameter of a near-Earth object to 140 meters and locate more than 90\% of them by 2020 \cite{ref-Harris} \cite{ref-Yeomans}. However, the new goal was not met even with a ten-fold increase in a NEO program budget \cite{ref-Martin}. As of Apr. 2018, more than 8,000 near-Earth asteroids of 140 meters and larger had been spotted while 17,000 such near-Earth asteroids were estimated to remain undetected. By 2019, the total number of discovered near-Earth asteroids of all sizes was more than 19,000 \cite{ref-Questions}.

In particular, the NASA funded Catalina Sky Survey (CSS) discovered over 1150 NEOs in years 2005 to 2007 \cite{ref-CSS}. In 2005, the Catalina Station (near Tucson, Arizona) based CSS became the most prolific NEO survey and surpassed Lincoln Near-Earth Asteroid Research (LINEAR) in the total number of NEOs and potentially hazardous asteroids discovered. As of 2020, the Catalina Sky Survey (CSS) has discovered 47\% of the total recorded NEOs, and the annual number has seen a steady growth \cite{ref-CSS2}. 

\subsubsection{Organization and Projects}

On the other hand, In June 2015, NASA and National Nuclear Security Administration of the U.S. Department of Energy officially formed joint cooperation \cite{ref-NYT}. Then in Jan. 2016, NASA officially announced the establishment of the Planetary Defense Coordination Office (PDCO) which aimed at identifying, tracking and warning about potentially hazardous NEOs that are larger than 30–50 meters in diameter and coordinating an effective emergency response, studying and developing mitigation technologies and techniques \cite{ref-PDCO}. PDCO has been involved in several key NASA missions, namely OSIRIS-REx, NEOWISE, NEO Surveyor, and DART which has been mentioned earlier. 

OSIRIS-REx, which represents Origins, Spectral Interpretation, Resource Identification, Security, Regolith Explorer, is a mission operated by NASA for the asteroid and solar system study and to obtain a sample of at least 60 g from an asteroid \cite{ref-Osiris}. OSIRIS-REx was launched in Sep. 2016 and rendezvoused with the asteroid 101955 Bennu in Dec. 2018 \cite{ref-Osiris2}. In Oct. 2020, it successfully touched down on Bennu and collected a sample with a mass of more than 60 g  and is expected to return to Earth in Sep. 2023 \cite{ref-Scientific}. 

Moreover, NEOWISE (Near-Earth Object Wide-field Infrared Survey Explorer) is a transferred and renamed mission re-activated from NASA infrared astronomy space telescope in Explorer program, WISE (Wide-field Infrared Survey Explorer) \cite[]{ref-WISE} \cite{ref-NEOWISE}. WISE was launched in Dec. 2009 and placed in hibernation mode in Feb. 2011, when it had already discovered thousands of minor planets, numerous star clusters, and Earth's first Trojan asteroid \cite{ref-Trojan}. Since the reboot of NEOWISE, NASA has been working with the Jet Propulsion Laboratory to investigate NEO threat scenarios and expects to find suitable solutions to mitigate the impact. 

Last but not least, NEO Surveyor, which was formerly called Near-Earth Object Camera (NEOCam) and then NEO Surveillance Mission, is also a space-based infrared telescope intended to survey the solar system for potentially hazardous asteroids larger than 140m \cite{ref-Surveyor}. The NEO Surveyor spacecraft will be working in the orbit of Sun-Earth $\text{L}_{\text{1}}$ \cite{ref-Surveyor}, \cite{ref-Smith}, the first Lagrange point of Sun-Earth, so that it can take a closer at the Sun and objects inside Earth's orbit \cite{ref-Surveyor} \cite{ref-SpaceNews}. As a successor to the NEOWISE mission, the NEO Surveyor mission was decided by PDCO to implement. In Jun. 2021, NASA authorized the mission to proceed to the preliminary design phase with the JPL leading the development \cite{ref-Surveyor}. In terms of \cite{ref-Surveyor2}, NEO Surveyor has used a 50-cm onboard telescope which is a bit larger than the 40-cm WISE telescope that has already successfully discovered 34,000 asteroids, including 135 NEOs. Its field of view is many times larger than WISE, enabling it to discover new NEOs with sizes as small as 30-50 m in diameter. \cite{ref-Surveyor2}. The detection improvement is largely accredited to new detector arrays (2,048 × 2,048 pixels and produce 82 Gb of data per day \cite{ref-Mainzer}) which have been modified to detect longer infrared wavelengths while being optimized for looking into cold space with excellent noise characteristics. Furthermore, the good infrared performance does not necessitate the use of cryogenic fluid refrigeration. The detector can be passively cooled to 30 K. Thus, there will not be a performance degradation due to running out of coolant \cite{ref-Mainzer}. 

\begin{figure}    %[H]
\centering
\subfigure[]{\includegraphics[width=4 cm]{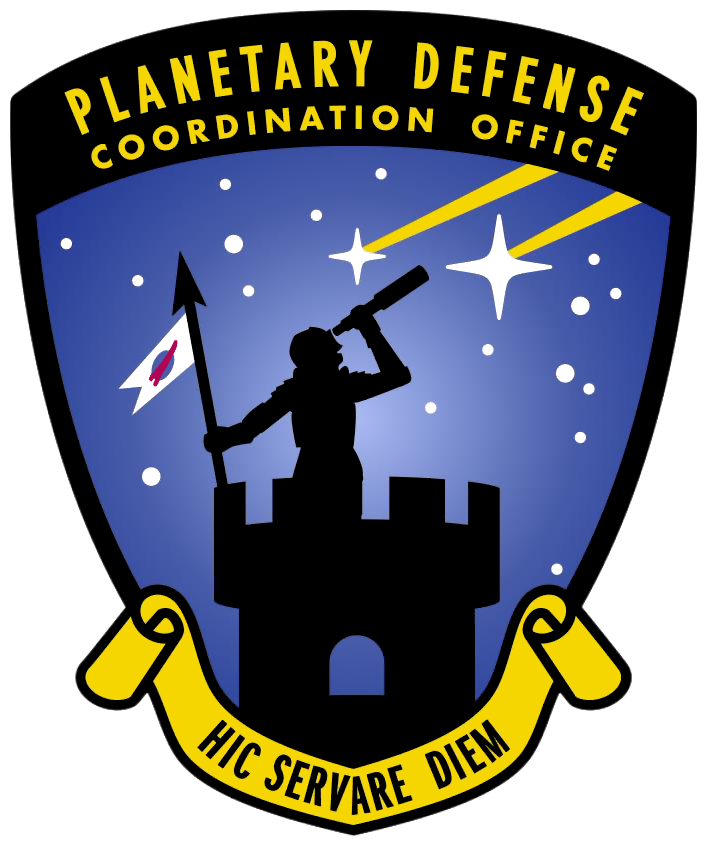}}
\subfigure[]{\includegraphics[width=4 cm]{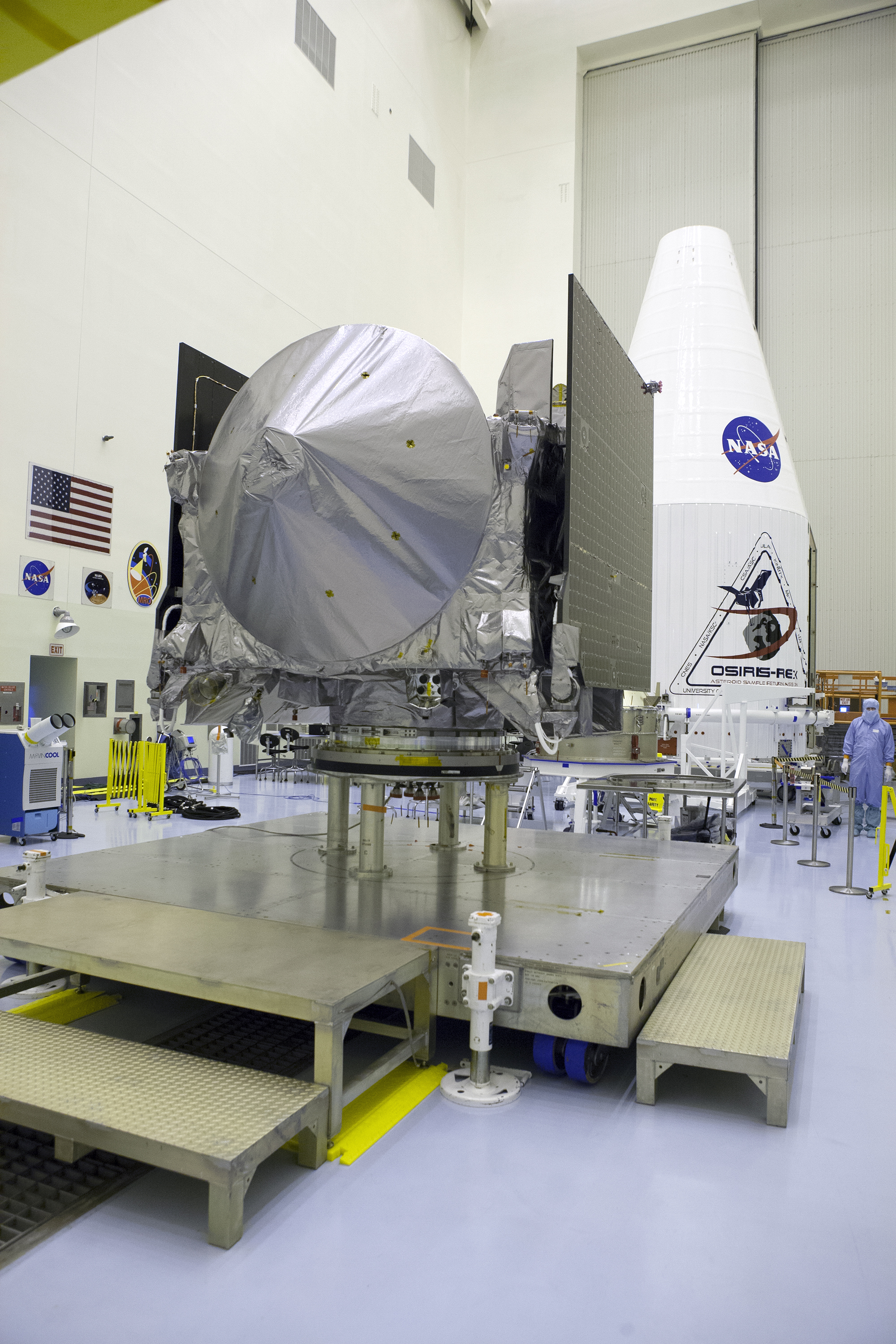}}\\
\subfigure[]{\includegraphics[width=6 cm]{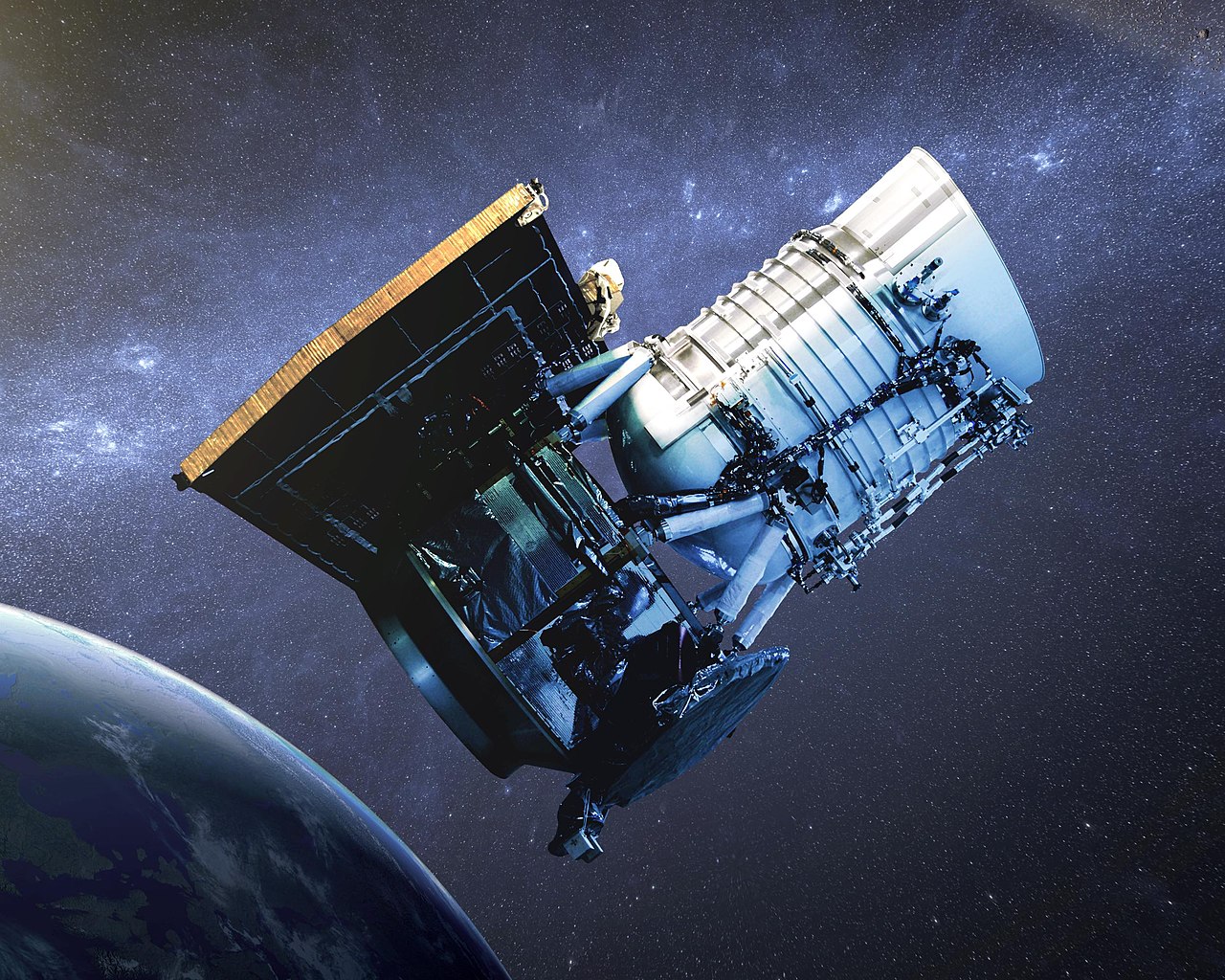}}
\subfigure[]{\includegraphics[width=6 cm]{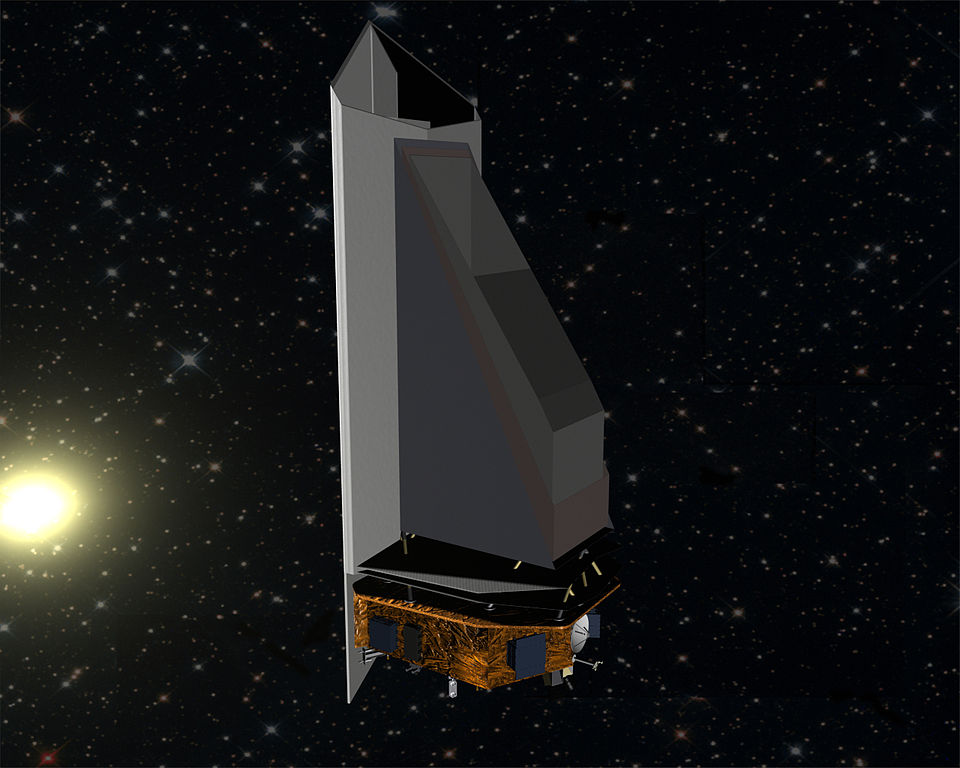}}
\caption{(\textbf{a}) The seal of NASA's Planetary Defense Coordination Office (PDCO) \cite{ref-Logo}. (\textbf{b}) OSIRIS-REx in launch configuration \cite{ref-NASA-Kennedy}. (\textbf{c}) Artist's concept of the Wide-field Infrared Survey Explorer (WISE) spacecraft in its orbit around Earth \cite{ref-NASA-WISE}. (\textbf{d}) Artist's illustration of the NEOCam (NEO Surveyor) space telescope \cite{ref-NEO-Cam}. \label{PDCO}}
\end{figure}  

Finally, the seal of the PDCO, OSIRIS-REx (in launch configuration), artist's depiction of the WISE (NEOWISE) and NEO Surveyor are illustrated in Figure~\ref{PDCO}, respectively.

%%%%%%%%%%%%%%%%%%%%%%%%%%%%%%%%%%%%%%%%%%
\section{Multi-planetary Defense: Live Long and Prosper}

\begin{figure*}%[H]
\centering
\centerline{\includegraphics[width = 26.6 pc]{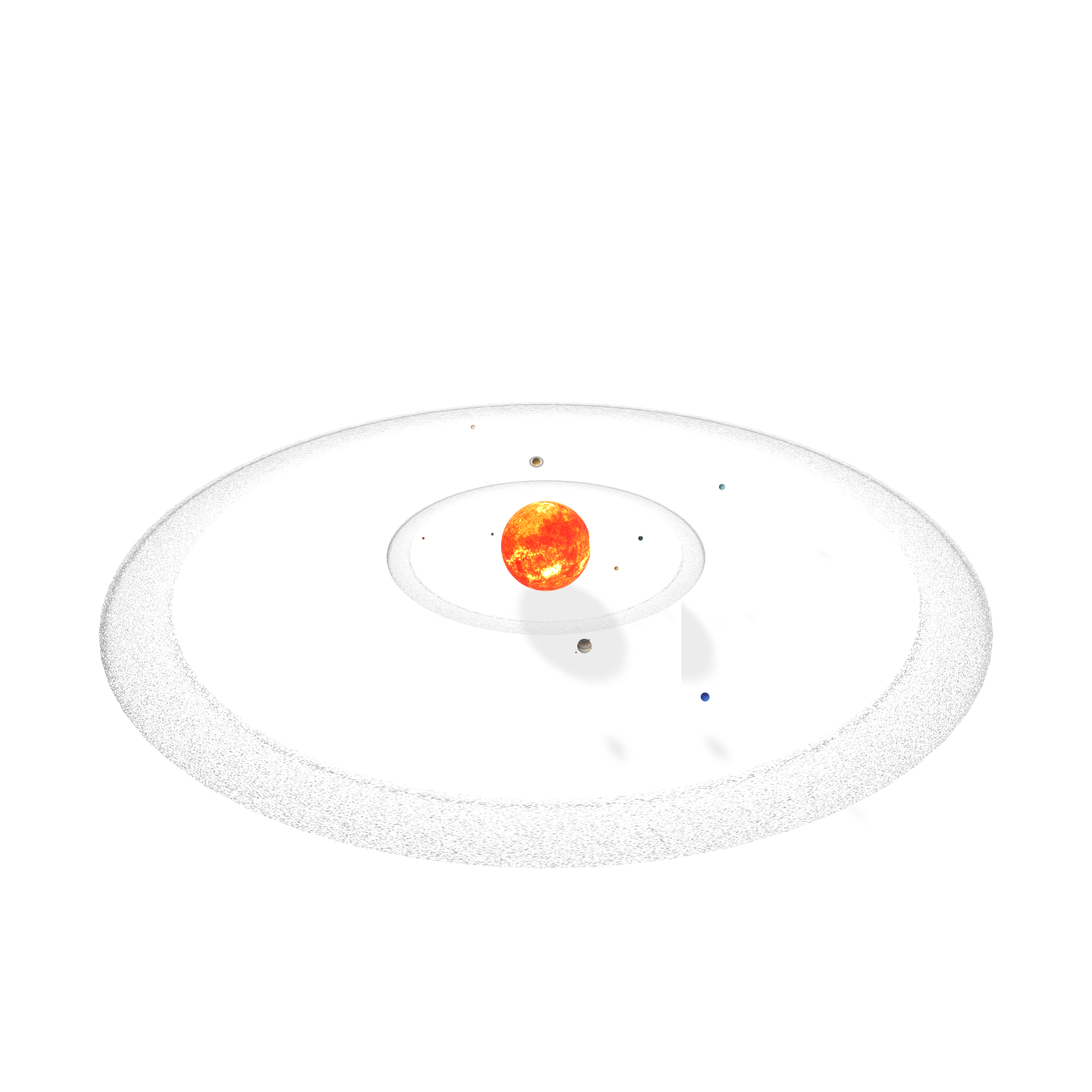}}
\caption{Illustration of the solar system with the Main Asteroid Belt and the Kuiper Belt.}\label{AsteroidBelts}
\end{figure*}

\subsection{Origins of Asteroids and Comets}

\subsubsection{The Main Asteroid Belt}

 The asteroid belt is also called the main asteroid belt (MAB) or main belt, a torus-shaped region in the inner Solar System located roughly between the orbits of Jupiter and Mars. It accommodates a large number of solid, irregularly shaped bodies, called asteroids or minor planets, as illustrated in Figure~\ref{AsteroidBelts}. With a total mass of approximately 4\% of Earth's Moon, the asteroid belt is the smallest and innermost circumstellar disc in the Solar System. Furthermore, about half of the mass of the asteroid belt stems from the four largest asteroids, namely, Ceres, Vesta, Pallas, and Hygiea \cite{ref-Williams}, and Ceres is the only dwarf planet with a mean diameter of 939.4 km \cite{ref-JPLCeres}. The main asteroid belt is believed to have formed from the very early-stage solar nebula as a group of planetesimals which are the smaller precursors of the protoplanets \cite{ref-Redd}. The gravitational perturbations from Jupiter, the most giant planet in the Solar System, pervaded the protoplanets with such high orbital energy that prevented them from accreting into a planet \cite{ref-Edgar}.  

The main asteroid belt is generally believed to be the source of most near-Earth objects. Authors in \cite{ref-Granvik} have performed extensive orbital integrations for a representative set of main-belt things to locate escape routes into the near-Earth space. Their proposed method can identify all essential regions for NEOs and provide a specific distribution of asteroids that will enter the NEO regions. Moreover, another investigation \cite{ref-Ormo} conducted by researchers from Europe and the United States discovered that approximately 470 million years ago, a 200 km sized asteroid was disrupted by a collision in the MAB, which generated many fragments into Earth's crossing orbits. During several millions of years following this one of the most significant cosmic catastrophic events, meteorite production and cratering rate had been significantly increased. For example, the 7.5 km wide Lockne crater in central Sweden is the consequence of the MAB event. Moreover, the authors provided evidence that the impact of a binary asteroid formed Lockne and its nearby companion, the 0.7 km diameter Målingen Crater.

Furthermore, the aforementioned Chelyabinsk asteroid is also considered to have an origin in the MAB. Based on the calculation of the pre-impact orbit of Chelyabinsk asteroid and the orbit of the asteroid 86039 (1999 NC43), researchers concluded that the pre-impact orbit is consistent with an origin in the MAB, most probably in the inner main belt near the $v_\text{6}$ secular resonance \cite{ref-Borovicka}.

\subsubsection{The Kuiper Belt}
With regard to comets which are cosmic snowballs of frozen gases, rock, and dust, they orbit the sun and have highly eccentric elliptical orbits with a wide range of orbital periods from several years to several millions of years. When passing close to the Sun, they are warmed up due to solar radiation and the solar wind and begin to release gases, which produces a visible atmosphere or coma, and sometimes also a tail. The tail can stretch away from the Sun for millions of miles \cite{ref-NASAComet}. As estimated, there are likely billions of comets orbiting the Sun in the Kuiper Belt and even more distant Oort Cloud. 

There are two major categories of comets in terms of their orbital periods. Short-period comets are believed to originate in the Kuiper Belt, which lies beyond the orbit of Neptune. As illustrated in Figure~\ref{AsteroidBelts}, the Kuiper Belt is a donut-shaped region of icy bodies in the outer Solar System \cite{ref-NASAKuiper}, extending from the orbit of Neptune at approximately 30 AU to 55 AU from the Sun \cite{ref-NASAKuiper}. The Kuiper Belt is similar to the MAB, but is way larger, around 20 times as wide and 20–200 times as massive \cite{ref-Delsanti}. It takes less than 200 years for short-period comets to orbit the Sun, usually the appearance of short-period comets is predictable since they have passed by before. The representative short-period comet is Halley's Comet, officially designated 1P/Halley and visible from Earth every 75–76 years.

Furthermore, long-period comets are believed to originate in the Oort cloud, which is a distant region of our solar system \cite{ref-NASAOort}. The Oort Cloud is believed to be a giant spherical shell surrounding the rest of the solar system. The inner edge is between 2,000 and 5,000 AU and the outer edge is 10,000 or even 100,000 AU (~1.58 light years), which is one-quarter to halfway between the Sun and the nearest neighboring star system, Alpha Centauri (a triple star system 4.35 light years from Earth) \cite{ref-NASAOort}. The Oort Cloud might contain billions of objects, or even trillions of objects larger than 1 km \cite{ref-Morbidelli}. Long-period comets move towards the Sun away from the Oort cloud due to gravitational perturbations caused by passing stars and the galactic tide. A typical example of long-period comets is Comet C/2013 A1 Siding Spring. It made a very close pass by Mars in 2014 and will not return to the inner Solar System for about 740,000 years \cite{ref-NASAOort}.

On the one hand, comets bombarding the young Earth about 4 billion years ago brought vast quantities of water. Organic molecules, such as polycyclic aromatic hydrocarbons \cite{ref-Clavin}, have been found in comets, which has led to speculation that comets may have brought the critical precursors of life or even life itself to Earth. On the other hand, comets may have caused many significant events on Earth. There have been debates on whether an asteroid or a comet was responsible for the Chixulub impact \cite{ref-Desch}. Similar discussions or uncertainties on the impactor have also been seen in the investigation of Tunguska event \cite{ref-Shoemaker1983} and Younger Dryas event \cite{ref-Temperature}. 

It is important to mention humanity's first direct observation of an extraterrestrial collision of Solar System objects, Comet Shoemaker–Levy 9's collision with Jupiter in July 1994. The comet, formally designated D/1993 F2, was discovered by astronomers Carolyn and Eugene Shoemaker and David Levy in 1993. By calculation, it may had been captured by Jupiter many years earlier and went through tidal breakup \cite{ref-Solem} in July 1992, eventually fragmented and collided with Jupiter in July 1994. This impact created a giant dark spot over 12,000 km (nearly one Earth diameter) across \cite{ref-AmericaSpace}, and was estimated to have released an energy equivalent to 6,000,000 megatons of TNT (600 times the world's nuclear arsenal) \cite{ref-Bruton}.

In addition, the asteroid/comet may rarely come from neither the Kuiper Belt nor the Oort Cloud. 'Oumuamua is the first known interstellar object detected passing through the Solar System. It was discovered by Pan-STARRS telescope located at Haleakalā Observatory in Hawaii on 19 October 2017 when it was 0.22 AU from Earth and already passed its closest point to the Sun approximately 40 days later \cite{ref-Oumuamua}. It was first classified as comet C/2017 U1 but reclassified again as asteroid A/2017 U1 due to the absence of a comet coma. Eventually, it was identified as 1I/2017 U1 because of its eligibility as an interstellar object. The authors conducted observation and verified a strongly hyperbolic trajectory in \cite{ref-Meech}. It has a hyperbolic excess velocity of 26.33 km/s (94,800 km/h), the speed relative to the Sun when traveling in interstellar space.

\subsection{Potentially Hazardous Objects and Detection}

\begin{figure*}   %[H]
\centering
\centerline{\includegraphics[width=26pc]{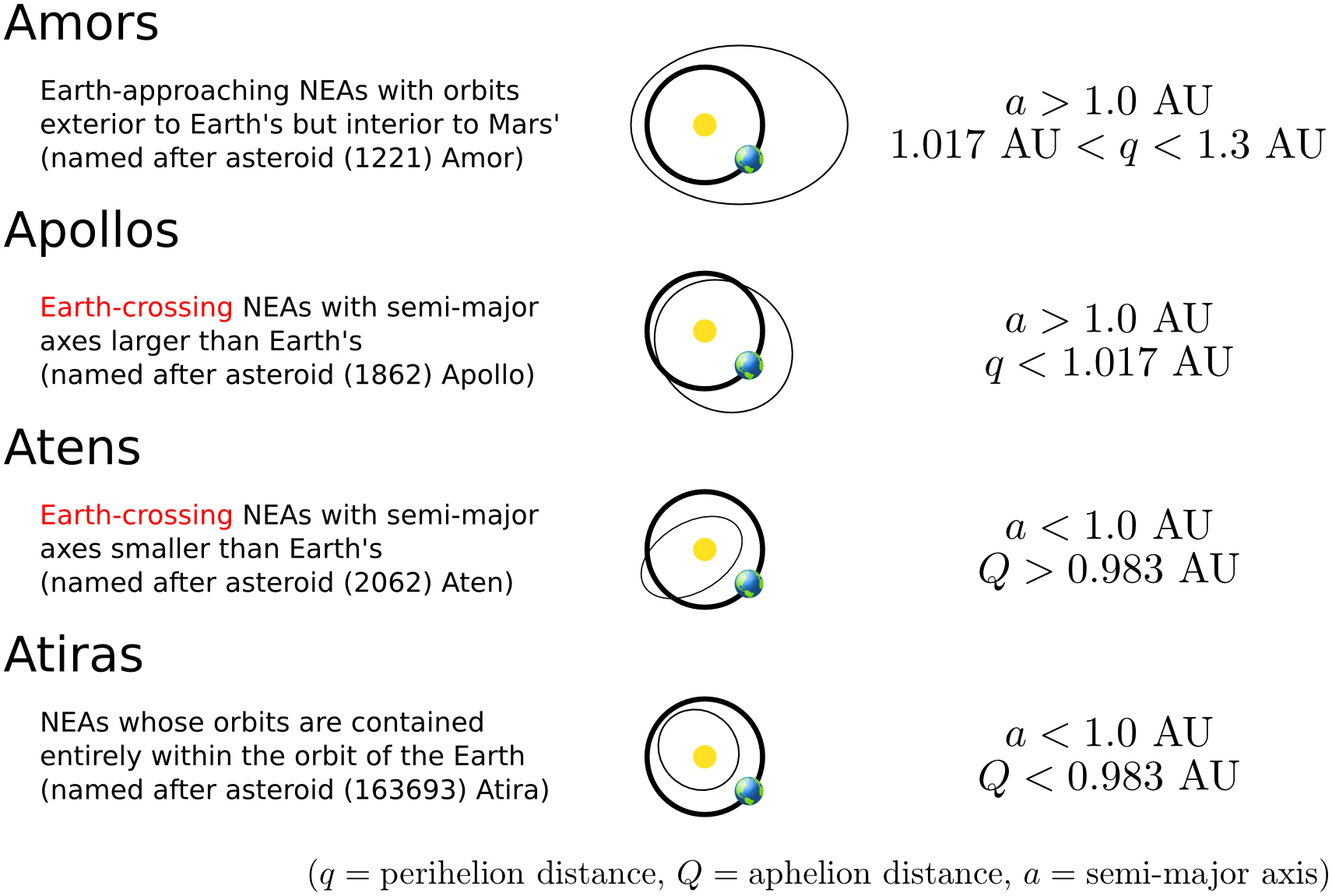}}
\caption{Four types of near-Earth asteroids known as Atira, Aten, Apollo and Amor \cite{ref-NEOs-Types}. \label{NEOs}}
\end{figure*}

According to \cite{ref-NEOs}, NEOs include asteroids and comets with perihelion distance $q$ less than 1.3 AU, while near-Earth comets (NECs) are specifically restricted to only short-period comets with an orbital period of fewer than 200 years. Near-Earth asteroids (NEAs) consist of the majority of NEOs, which are categorized into four types, namely, Atira, Aten, Apollo, and Amor, according to their perihelion distance (q), aphelion distance (Q) and their semi-major axes (a). The definitions and relations of these four types are given and illustrated in Figure~\ref{NEOs}.  

Furthermore, potentially hazardous objects (PHOs) are currently defined in terms of parameters that evaluate the asteroid's potential to make threatening close approaches to Earth \cite{ref-NEOs}. For example, all objects with an Earth minimum orbit intersection distance (MOID) of 0.05 AU (7,480,000 km or 19.5 lunar distances) or less and an absolute magnitude (H) of 22.0 or less (equivalent to a diameter of at least 140 m) are considered PHOs. This dimension (140 m in diameter) is large enough to cause unprecedented regional devastation to human settlements in the case of a land impact or a very dangerous tsunami in the case of an ocean impact. 

\begin{figure*}    %[H]
\centering
\centerline{\includegraphics[width=26pc]{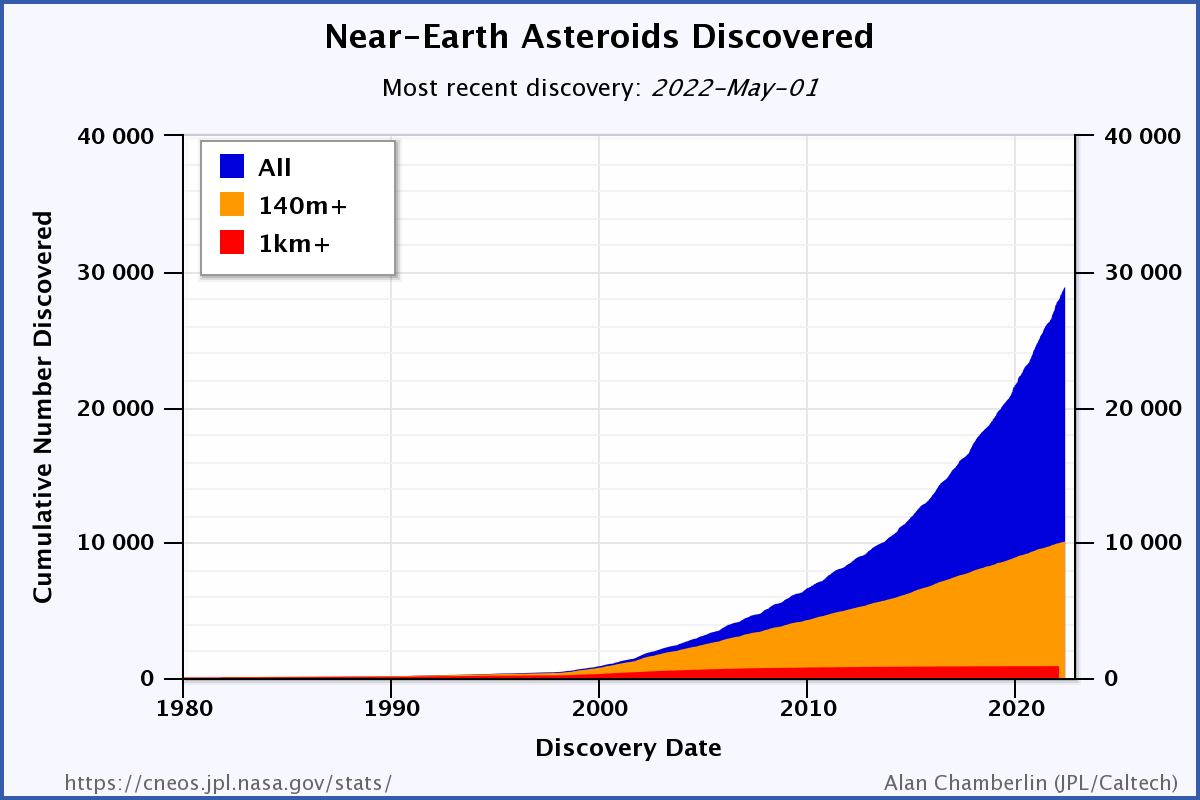}}
\caption{Near-Earth asteroids discovered as of May 01, 2022 \cite{ref-PHOs}. \label{NEOs_Stat}}
\end{figure*}

\begin{figure*}    %[H]
\centering
\centerline{\includegraphics[width=24pc]{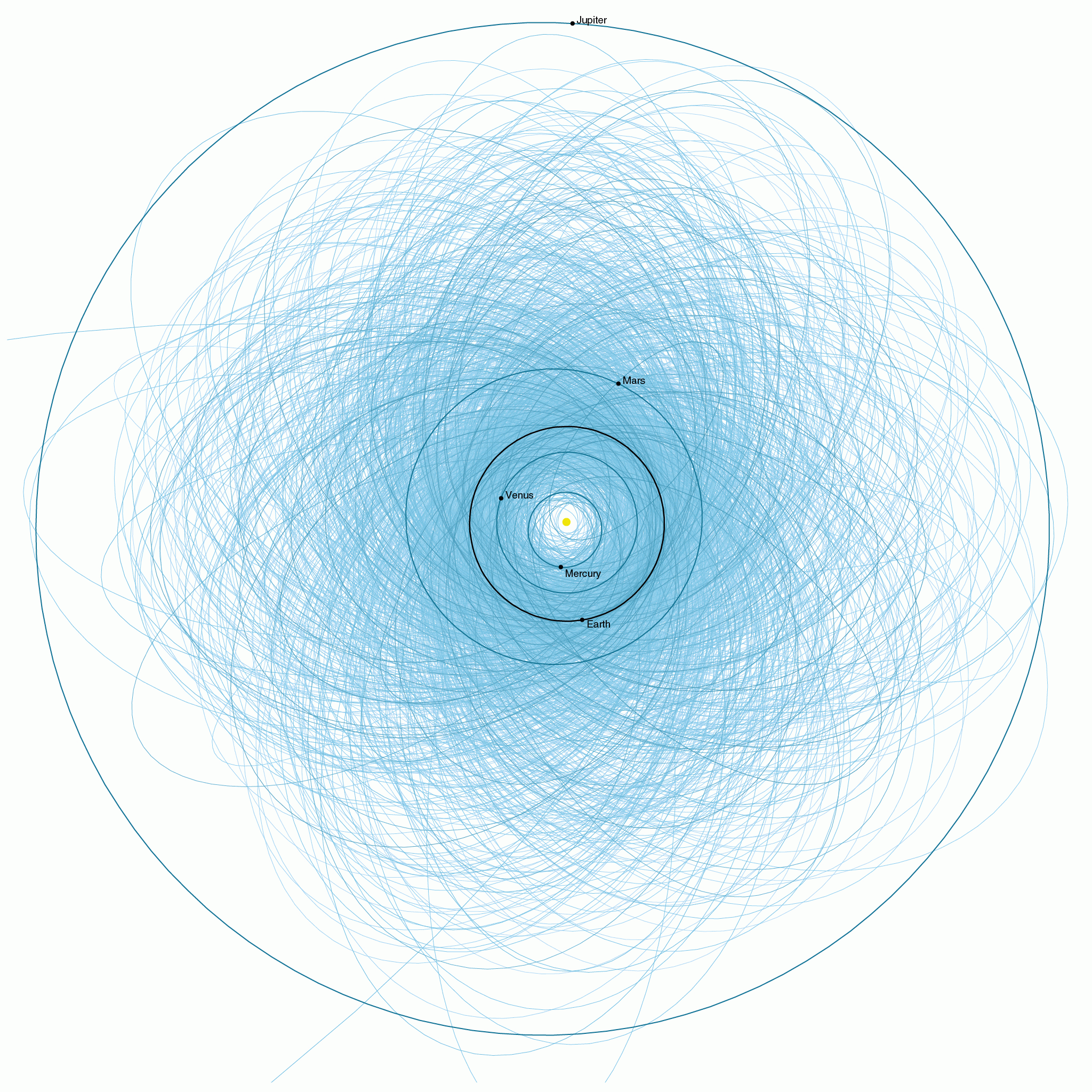}}
\caption{Plot of orbits of known potentially hazardous asteroids with sizes over 140 metres as of early 2013 \cite{ref-PHOs-Orbits}. \label{PHA}}
\end{figure*}

As of May 01, 2022, there are a total of 28,874 NEOs discovered and surveyed \cite{ref-PHOs}. Only 117 (0.4\%) of them are NECs, and the remaining are all NEAs. Furthermore, as illustrated in Figure~\ref{NEOs_Stat}, 10070 of all NEAs are NEA-140m (with diameters 140 meters and larger), while 877 are NEA-km (with diameters 1 km and larger). Moreover, 2262 and 156 of all NEOs are categorized into PHA and PHA-km, respectively \cite{ref-PHOs}. Eventually, the orbits of all known PHOs can be plotted for real-time tracking, such as in Figure~\ref{PHA}.

In terms of theoretical estimations, impact events caused by PHOs in the 140-meter and larger dimension may occur on average around once at least every 10,000 years, while a meteoroid the size of a football field hits Earth every 2,000 years or so \cite{ref-NASAFacts}. However, due to many uncertainties and factors, it is hard to predict when and where the impact will happen or if humanity will be ready to handle it. In addition, an impact event that is caused by PHOs smaller than 140 meters and occurs in a shorter time period cannot be ignored since the aforementioned Tunguska event, and Chelyabinsk event were both caused by meteorites much smaller than 140 meters. In fact, authors in \cite{ref-Earth} implied that ‘Tunguska’ impact rate on Earth is one event every 300 years and probably rather more frequent, which is much shorter than the current preferred value of 2,000–3,000 year.     

\subsection{State of the Art Space Objects Detection}

\subsubsection{Detection Capability According to Object Dimension}
Prior to proposing any possible evolution plan for space object detection, it is noteworthy to review and conclude the performance of existing ones. The above-mentioned Spaceguard survey activities involved several entities such as Lincoln Near-Earth Asteroid Research, Spacewatch, Near-Earth Asteroid Tracking (NEAT), Lowell Observatory Near-Earth-Object Search (LONEOS), Catalina Sky Survey, Campo Imperatore Near-Earth Object Survey (CINEOS), Japanese Spaceguard Association, Asiago-DLR Asteroid Survey (ADAS) and Near-Earth Object WISE. Thanks to the global joint efforts, most NEAs larger than 1 km in diameter (NEAs-km) have been surveyed and tracked. As of May. 02, 2022, 877 NEAs-km have been discovered, which accounts for 95.3\% of an estimated total of about 920 NEAs-km \cite{ref-Tricarico}.

However, for NEAs with absolute magnitude $17.75 < H < 22.75$ (corresponding to a diameter larger than 100 m but smaller than 1 km), 12,058 of them have been surveyed as of May. 01, 2022, which however averagely takes up only 15.8\% of the estimated total ($(7\pm2) \times 10^4$) \cite{ref-Tricarico}. Based on the current detection pace, as shown in Figure~\ref{NEOs_Stat}, it might take tens of more years to locate all these NEAs. Moreover, an even higher proportion of asteroids smaller than 100 m is not yet located, which could take more resources and time to accomplish. This can also be interpreted that the current detection capability for asteroids of some specific dimension is still relatively limited, which could weaken the success rate of early warning and slow the reaction effort and mitigation deployment when a real threat appears. 

\subsubsection{Detection and Early Warning}
Generally speaking, the current detection capability of NEOs is highly related to both the dimension and perihelio distances of NEOs. Theoretically, the smaller a NEO's dimension or perihelio is, the harder it can be detected. The current detection system and detection range are mainly intended for near-Earth asteroids/comets with a maximum perihelio distance of 1.3 AU. It will be more challenging to survey and locate non-NEO asteroids/comets such as in the MAB, Kuiper Belt, or even further regions of the solar system. However, these asteroids/comets can potentially become NEOs when a collision or gravitational disturbance flings them inward \cite{ref-JPLWise}. Authors in \cite{ref-Granvik} have conducted an in-depth analysis of possible escape routes for asteroids to enter the near-Earth space.   

On the other hand, there are possibly billions of comets that are orbiting the Sun and located in the Kuiper Belt and even more distant Oort Cloud. However, only 3,743 of them so far have been discovered according to NASA \cite{ref-NASAComet}. The comets' orbits could be potentially altered due to gravitational disturbance, solar flare burst, collision, etc. As a result, it would not be a rare scenario for a comet enters near-Earth space at high speed. 'Oumuamua is a typical case, and it was only 0.22 AU away from Earth when we discovered it \cite{ref-Oumuamua}. Furthermore, it has been reported by NASA astronomers that 6 months' warning is not enough, and 5 to 10 years of preparation may be needed to stop an asteroid from hitting Earth, based on the simulated exercise conducted by the 2021 Planetary Defense Conference \cite{ref-PDC2021}. Consequently, it will be critical and necessary to enhance the detection capability so that more preparation time could be obtained for humanity.

\begin{figure*}%[H]
\centering
\centerline{\includegraphics[width=36pc]{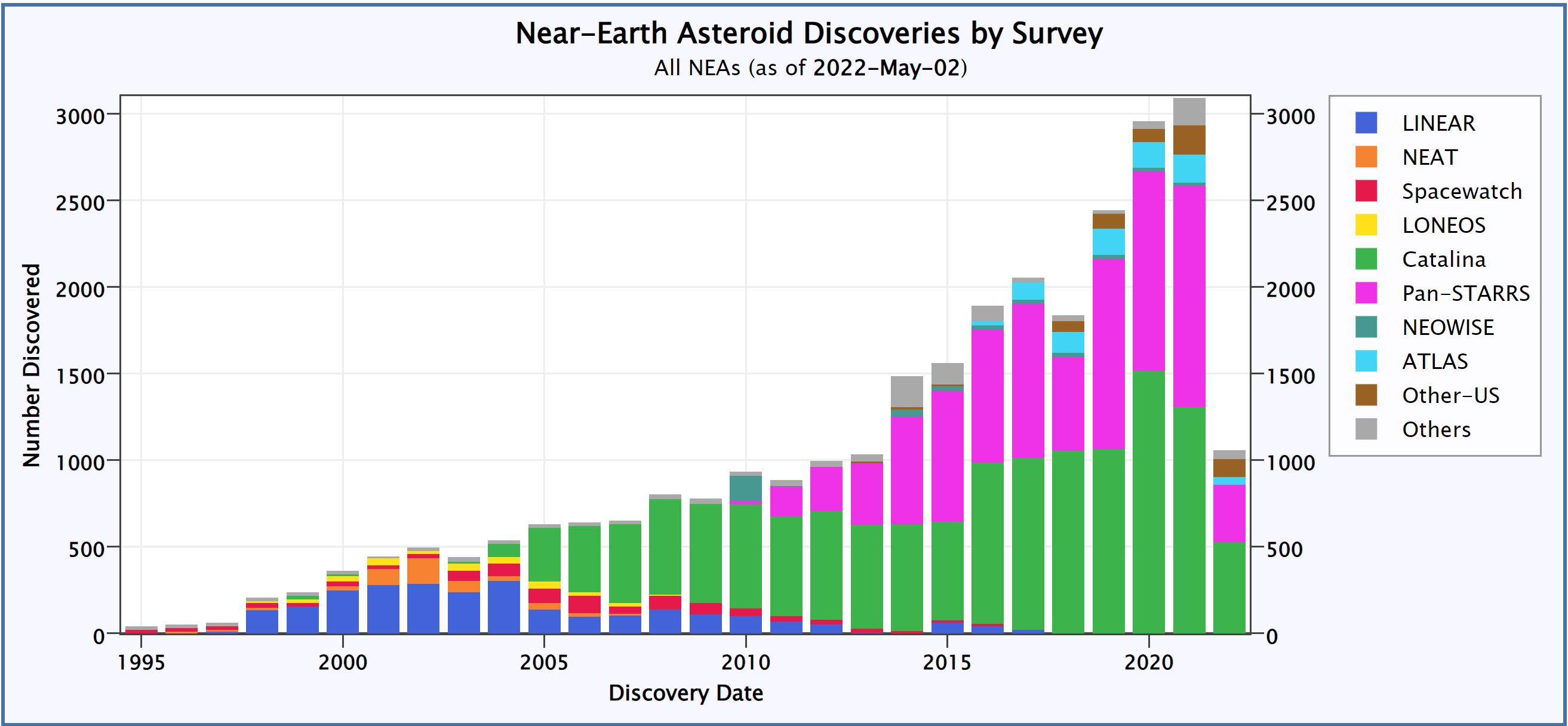}}
\caption{Plot of the number of NEA discoveries per year, by survey \cite{ref-PHOs}. \label{NEASurvey}}
\end{figure*}

\subsubsection{On-Ground Observatories}
One of the possible and feasible strategies is to deploy more survey stations dedicated to discovering more asteroids/comets, which are not only categorized as NEOs but also are located in more distant regions. When reviewing the current near-Earth asteroid discoveries by survey, as illustrated in Figure \ref{NEASurvey}, a majority of NEA discoveries have been made by Arizona-based Catalina and Hawaii-based Pan-STARRS since 2011. The Catalina Sky Survey uses three telescopes: a 1.5 meter f/1.6 Cassegrain reflector telescope on the summit of Mount Lemmon (around 2791 m), a 68 cm f/ 1.7 Schmidt telescope near Mount Bigelow, and a 1-meter f/2.6 follow-up telescope also on Mount Lemmon. Furthermore, the Panoramic Survey Telescope and Rapid Response System (Pan-STARRS) is located at Haleakalā Observatory, Hawaii, US, and consists of two 1.8 m Ritchey–Chrétien telescopes on the summit of Haleakalā (also known as East Maui Volcano) which has an elevation of 3,055 meters. The construction of both telescopes was funded by the U.S. Air Force and the NASA Near-Earth Object Observation Program. Both locations offer ideal observation conditions; for example, the air pollution and light pollution are minimal, the air density is significantly lower, the annual precipitation is extremely low, and the temperature fluctuation through the year is also very small. Furthermore, in astronomy, seeing refers to the degradation of the image of an astronomical object due to turbulent airflows in the atmosphere of Earth, and it may cause the blurring, twinkling or variable distortion of the image. The Haleakalā observatory can also provide the best condition in terms of seeing. Another place providing such conditions is Roque de los Muchachos Observatory, located on the island of La Palma in the Canary Islands, Spain.

Therefore, finding ideal locations for ground observatories is challenging since many unique conditions are expected to be fulfilled. Moreover, when looking at the bigger picture, the ground observatories in the Northern Hemisphere or the Southern Hemisphere can only observe some portion of the entire sky. Joint efforts should be synchronized to thoroughly survey the entire sky in order not to miss any suspicious near-Earth object. In fact, the ideal place for founding observatories in the Southern Hemisphere is in the Atacama Desert, which is a desert plateau in northern Chile, covering a 1,600 km strip of land on the Pacific coast, west of the Andes Mountains. The Atacama Desert has a high elevation and is the driest nonpolar desert globally, receiving less precipitation than the polar deserts. Due to its unique geological and weather conditions, some of its regions have been used for astronomical observatories and experimentation sites on Earth for Mars expedition simulations. The European Southern Observatory (ESO) operates three major observatories in the Atacama, which are known as, La Silla Observatory, Paranal Observatory, Llano de Chajnantor Observatory, and ESO is currently building a fourth one, Cerro Armazones Observatory, site of the future Extremely Large Telescope (ELT). The Llano de Chajnantor Observatory includes the famous Atacama Large Millimeter/submillimeter Array (ALMA), which is an astronomical interferometer of 66 radio telescopes and observes electromagnetic radiation at millimeter and submillimeter wavelengths. The array was constructed on the 5,000 m elevation, which is crucial to reducing noise and decreasing signal attenuation due to Earth's atmosphere. However, these telescopes are not mainly dedicated to the NEO survey. 

On the other hand, on-ground telescopes for the NEO survey mainly use radio frequencies and visible light. Only a few of them are infrared-based, such as NASA Infrared Telescope Facility (NASA IRTF). Infrared telescopes hold some advantages, one of which is, for example, when observing in the near-infrared, the dust is transparent to it. Therefore, this explains why an optical telescope would be unable to see a star hidden in dust, whereas one working in the near-infrared would be able to detect its emission. Another advantage is that relatively cold objects invisible to optical telescopes become visible in the infrared. Interstellar gas, dust discs, asteroids, and brown dwarfs are all examples of objects that are too cold to shine in visible wavelengths but become conspicuous when viewed in the infrared \cite{ref-ESO_Why}. Consequently, an infrared facilitated survey is beneficiary for detecting asteroids/comets, which are usually cold and hard to be discovered. However, most infrared telescopes such as Herschel, NEOWISE, JWST, are placed in space to completely eliminate the interference from the Earth's atmosphere. Furthermore, cryogenic technology will be needed with spaceborne infrared detection since the low temperature can effectively suppress the detection of dark currents and background noise \cite{ref-Cryogenic}.  

Moreover, from the radio astronomy perspective, powerful ground-based radio telescopes can effectively detect NEOs. For example, the California-based Goldstone Deep Space Communications Complex (GDSCC) made a historical observation of the 1,000th near-Earth asteroid \cite{ref-NASA1000} on Aug. 22, 2021, on its Deep Space Station (DSS) 14: "Mars" where an enormous Cassegrain antenna of 70-m diameter is installed. When created by the JPL to support the Pioneer program of deep space exploration probes, the location of the Goldstone complex was chosen to be in the Mojave Desert, distant from radio interference. The GDSCC was mainly intended for the vital two-way communications link that tracks and controls interplanetary spacecraft. However, it can also be used as high-sensitivity radio telescopes for astronomical research, such as radar mapping planets, asteroids, and comets.

Furthermore, the advances in astronomical spectroscopy also provide better measurement of the spectrum of electromagnetic radiation from celestial bodies. Scientists could detect molecules of interesting materials on celestial bodies and better understand the solar system's evolution. For example, a semiconfocal cavity coupled pulse echo spectrometer system designed in 65-nm CMOS is able to detect Nitrous Oxide (N$_2$O) that re-emits at 100.4917 GHz \cite{ref-Tang2017}. Furthermore, a 180-GHz pulsed CMOS transmitter with a high output power of 0.6 mW is presented for emission-based molecular detection \cite{ref-Nemchick}. It is predicted by Dr. Tang that the emission spectroscopy will play a crucial role in analyzing a plume or gas emission of a celestial body and being used in radio-astronomy looking at distant stellar objects. Therefore, emission spectroscopy can be also utilized to conduct precise spectroscopic analysis of the asteroids that are currently classified into three major types according to their spectra. In the former Tholen classification, the C-types are made of carbonaceous material, S-types consist mainly of silicates, and X-types are metallic. Then in 2002, the Tholen classification was upgraded into the SMASS classification, expanding the number of categories from 14 to 26 \cite{ref-Bus}.

\subsection{Evolving the Networks of Survey Stations}
We can summarize several key points from previous investigations as follows:
\begin{itemize}

 \item \emph{\textbf{First}}, due to many types of random events and their consequences, asteroids, and comets can escape or change their former orbits and enter the near-Earth space, which increases the probability of their impact on Earth. The current asteroids/comets detection is focused on the NEO category, which may limit the preparation time allowed for humanity. Although more than 96\% of total estimated large asteroids (with a diameter of 1 km and more) in the near-Earth space have been surveyed so far, the proportion of NEAs (with a diameter between 100 m and 1000 m) is still very low.  

 \item \emph{\textbf{Second}}, the detection capability of on-ground telescopes highly depends on their geological and meteorological conditions. Only several specific locations on Earth can fulfill such challenging requirements for high-quality observation. Moreover, most on-ground observatories operate in visible light and radio frequencies. Just a few ground-based telescopes work in infrared wavelengths, and many infrared telescopes are space-based to minimize the Earth's interference and background noise. 
 
 \item \emph{\textbf{Third}}, the ground-based observatories usually are powerful with abundant facilities and local technical support, while space-based ones normally can only depend on the solar array and remote debugging. On the other hand, space-based telescopes have obvious advantages over the on-ground ones to survey the sky in infrared wavelengths, thus playing a crucial role in identifying and locating NEOs. For example, most asteroids are black, and small ones are difficult to see in the blackness of outer space with an optical telescope. Still, a telescope operating at infrared wavelengths is sensitive to asteroids' surfaces warmed by the Sun.  
 
\end{itemize}

Based on the features and characteristics of asteroids/comets and related detection technologies, building and deploying more space-based survey stations with multiple techniques can accelerate the discovery of all potentially hazardous objects (PHOs). Furthermore, with minimal interference from Earth, space-based telescopes and spacecraft can more effectively survey the universe, with access to the full sky when deployed properly. In addition, the space-based survey stations are connected with on-ground stations and resources to form a more comprehensive and reliable network for PHOs detection. For the next, we can propose a framework for the solar communication and defense networks (SCADN) which consists of the following critical components:

\begin{enumerate}

\item {A large number of survey stations (on spacecraft) will be deployed into various orbits across the entire solar system to form an enormous internet of spacecraft (IoS) networks. As humans are supposed to become a multi-planetary species after extra-terrestrialization, the colonization of other celestial bodies could also fall victim to asteroid/comets impact. For example, due to the lack of a thick atmosphere as on Earth, Mars could be more vulnerable to the impact of near-Mars objects (NMO). Moreover, Jupiter and Saturn, the gas giants which are 318 times and 95 times as massive as Earth, can affect and steer some asteroids away from Earth, thus providing some protection to humanity. However, Jupiter and Saturn attract asteroids/comets to their regions and increase the impact probability on their moons including Europa \cite{ref-Europa}, Titan and Enceladus \cite{ref-Enceladus} which are particularly interesting and may be possible for founding future human colonization. Similarly, some moons of ice giants, Uranus and Neptune, also can serve as humanity's colonization and outpost for even further expeditions in the universe. Consequently, deploying survey stations and spacecraft into regions/orbits of Mars, Jupiter, Saturn, Jupiter, Saturn, Uranus, and Neptune is also crucial.}

\begin{figure*} %[H]
\centering
\centerline{\includegraphics[width=36pc]{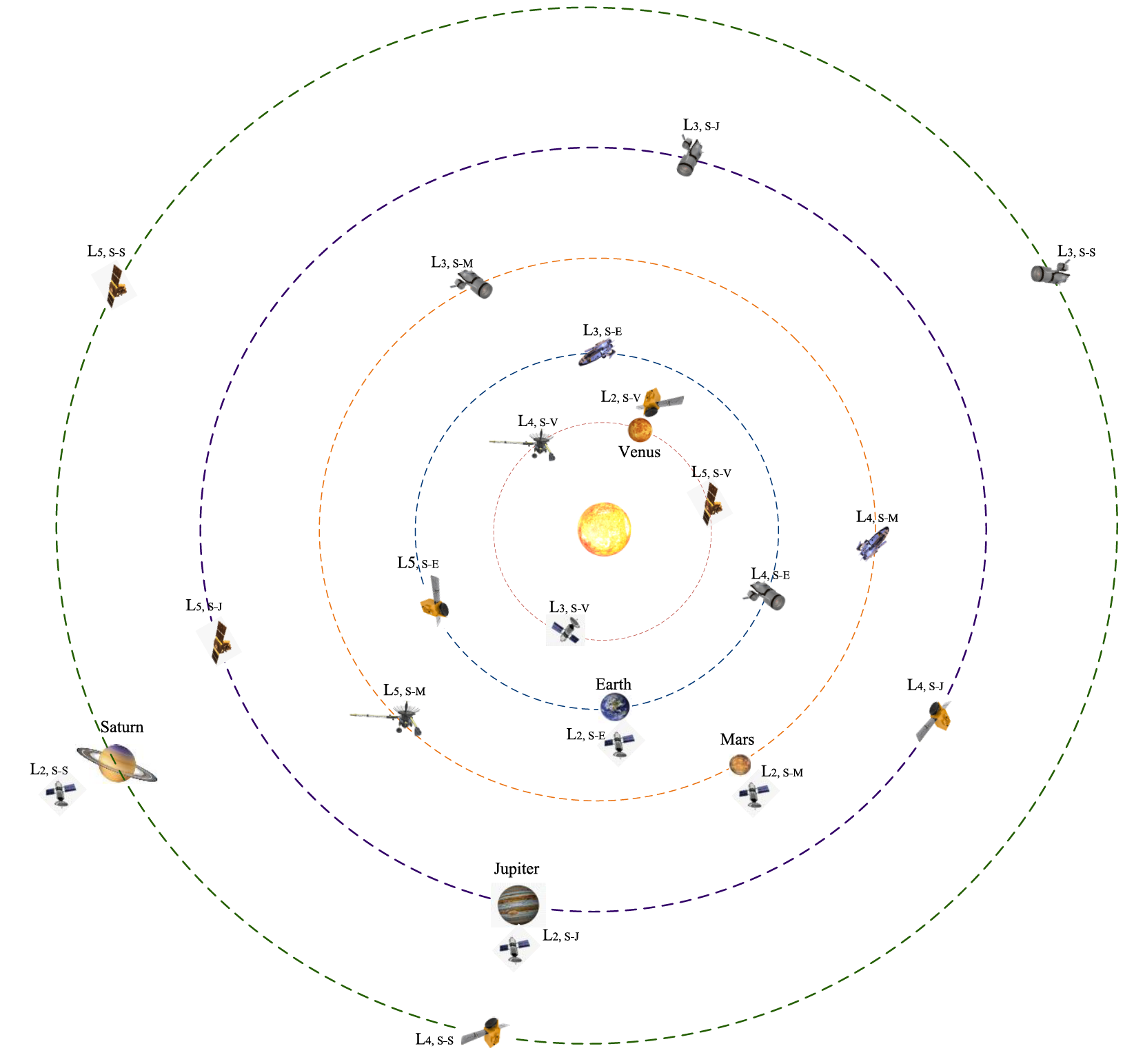}}
\caption{An exemplary illustration of survey stations and spacecraft deployed under the SCADN framework, in particular, the survey stations/spacecraft are deployed at Lagrange points of the solar planets (dimensions of orbits, planets and spacecraft are not scaled). \label{SCADN}}
\end{figure*}

\item {As illustrated in Figure \ref{SCADN}, for each planet in the solar system, multiple survey stations/spacecraft are deployed in the planet's orbit. In the SCADN, the specific locations for accommodating these spacecraft can be on Lagrange points where the gravitational forces of the two large bodies and the centrifugal force balance each other so that spacecraft only require minimal orbital corrections. Take Earth, for example; multiple survey stations/spacecraft can be deployed on Earth-Sun Lagrange points, L$_\text{2, S-E}$, L$_\text{3, S-E}$, L$_\text{4, S-E}$, and L$_\text{5, S-E}$, respectively. There are several features enabled by adopting such a strategy. First, L$_\text{2, S-E}$, which has a distance of around 0.01 AU from Earth, can enable a high-performance survey (with minimal interference from the Sun) to the space in the direction away from the Sun but a reliable communication with Earth. Second, L$_\text{3, S-E}$, L$_\text{4, S-E}$, and L$_\text{5, S-E}$ can help survey the space more thoroughly and completely than on/near-Earth stations. For instance, the object within a specific elongation from the Sun, which is a region of the sky giving no access to ground-based telescopes, will be discovered by stations deployed in these locations. Therefore, the probability of failed detection and alarming of incoming objects from outer space (such as Chelyabinsk meteorite) will be minimized.}

\item {Furthermore, as shown in the exemplary illustration, the Lagrange points of other planets such as Venus, Mars, Jupiter, Saturn, will also be deployed with survey stations/spacecraft. For example, Venus has an average distance of 0.72 AU from the Sun and orbits the Sun faster than other outer planets. Therefore, deploying survey stations/spacecraft into the four Lagrange points of Venus can also help survey the potentially hazardous objects approaching other outer planets (of Venus) more effectively. Since the space increases over the growth of orbits, deploying more survey stations/spacecraft for planets of interest using this method will cover more space and further minimize the failure detection and warning. It is noteworthy that Figure~\ref{SCADN} is just an exemplary case, Mercury, Uranus, and Neptune are not shown for simplicity. All survey stations and spacecraft are inter-connected and can collaborate on a series of survey tasks to realize a better overall performance gain. On each Lagrange point, there may be more than one survey stations/spacecraft needed to fulfill the specific requirement. Eventually, a very large sphere of at least more than 30 AU in the radius can be covered under the SCADN so that many more space objects that previously could not be located/tracked by humans could be fully available in the database. Moreover, these survey stations/spacecraft could serve for other space exploration and scientific experiment tasks, thanks to their unique celestial coordinates. }

\item {All survey stations and spacecraft are expected to communicate and connect with each other using multiple state-of-the-art communications and networking technologies. Some of the unprecedented challenges for wireless communications are, the very long distances over which the wireless signals need to travel, and the solar flare that can seriously interrupt the radio communication. Also, the meteorological conditions on Earth and other celestial bodies can lead to degradation in wireless communications. Although the low data rate communication over very long distances in the deep space is proven to be working well (such as voyager 1 and 2 that are currently around 155.6 AU and 129.9 AU from the Sun), high data rate one can still be a challenging one even today. Another challenge is the significant latency. With 1 AU distance meaning 499 seconds for light to travel in the free space, the distance between Saturn's L$_\text{3, S-S}$ and Saturn itself is 19 AU or 9481 seconds (around 158 minutes) for light to fly. The communication latency between Neptune and its L$_\text{3, S-N}$ is around 500 minutes. Moreover, the latency between Earth and Lagrange points of different planets varies over time.}

\item {Considering the challenges of wireless communications over extremely long propagation, particularly the very large latency, using Earth as the only routing point to store, exchange, and process data and telemetry commands from survey stations/spacecraft may lead to a very long latency and low efficiency. There are several potential solutions to this dilemma. For example, it would be beneficiary to enable artificial intelligence (AI) and edge computing on the survey stations/spacecraft so that they could process the image and data extracted from the space and determine if any detection falls into interest or any further resource (other survey stations/spacecraft and computing power) is needed to assist in the task. Moreover, any communication interruption due to interference could be mitigated when using some of the available survey stations/spacecraft as relays. Eventually, such an AI and edge computing assisted, cell-free architect can improve the overall system efficiency of wireless communications in deep space and identify the objects of interest.}      

\end{enumerate}

\subsection{Equipping The SCADN With Mitigation Technologies}
On top of the initial development and deployment of the SCADN, some further proactive strategies of mitigation can be utilized to reduce the probability of catastrophic consequences of the impact by the space objects. Conventionally speaking, the mitigation techniques, or collision avoidance techniques, are developed based on metrics such as technology readiness, failure risks, operation feasibility, performance, and cost \cite{ref-Canavan2}. There have been various methods proposed to change the course of an asteroid or comet \cite{ref-Hall}, which can be categorized by various types of attributes such as the type of mitigation, approach strategy and energy source, as summarized and concluded in Table~\ref{table:mitigation}.  

As observed, the energy source of the nuclear explosive device can provide the most effective approach to cope with PHOs of various sizes for either short-notice or long-notice threats, particularly against solid objects. However, many NEOs are believed to be loosely held together by gravity as "flying rubble piles" and thus cannot be effectively handled by the method of kinetic impactors or nuclear explosive devices. The indirect methods such as gravity tractor, focused solar energy, laser ablation, ion beam shepherd might take more time to alter the PHO's trajectory but require early development and deployment, such as traveling to the PHO's proximity in advance, for the space rendezvous. Furthermore, the PHO deflection means of gravity tractor, focused solar energy, the deployment procedure of the DART mission, and are illustrated in the following figure, respectively. 

In addition, some other proposed deflection methods include wrapping the PHO in a sheet of reflective plastic such as aluminized polyethylene terephthalate (PET) as a solar sail; dusting the PHO with titanium dioxide (white) to alter its trajectory via increasing the reflected radiation pressure or with black to alter its trajectory via the Yarkovsky effect; attaching a mass driver on the PHO to eject material into space in order to give the PHO a slow steady push and decrease its mass; deploying coherent digging/mining array multiple 1-ton flat tractors able to dig and expel PHO's soil mass as a coherent fountain array \cite{ref-HowWhereWhen}; attaching a tether and ballast mass to the PHO to alter its mass center and trajectory; using magnetic flux compression to magnetically brake and or capture a PHO that contains a high percentage of meteoric iron by deploying a wide coil of wire in its orbital path \cite{ref-Durda}.  

The multi-layered architecture of the SCADN network can detect and mitigate the PHO more effectively. On top of a deep understanding of the characteristics of mainstream PHO mitigation technologies, the SCADN framework is equipped with various types of mitigation strategies with several key points unfolded as follows. 

\begin{figure} % [H]
\centering
\subfigure[]{\includegraphics[width=6.5 cm]{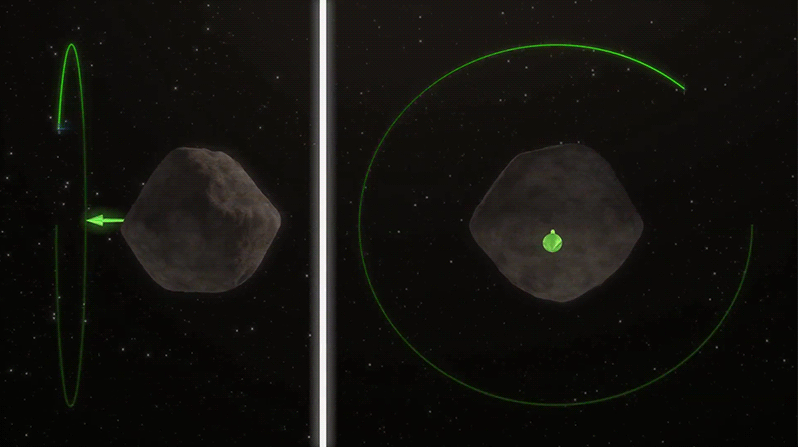}}
\subfigure[]{\includegraphics[width=6.5 cm]{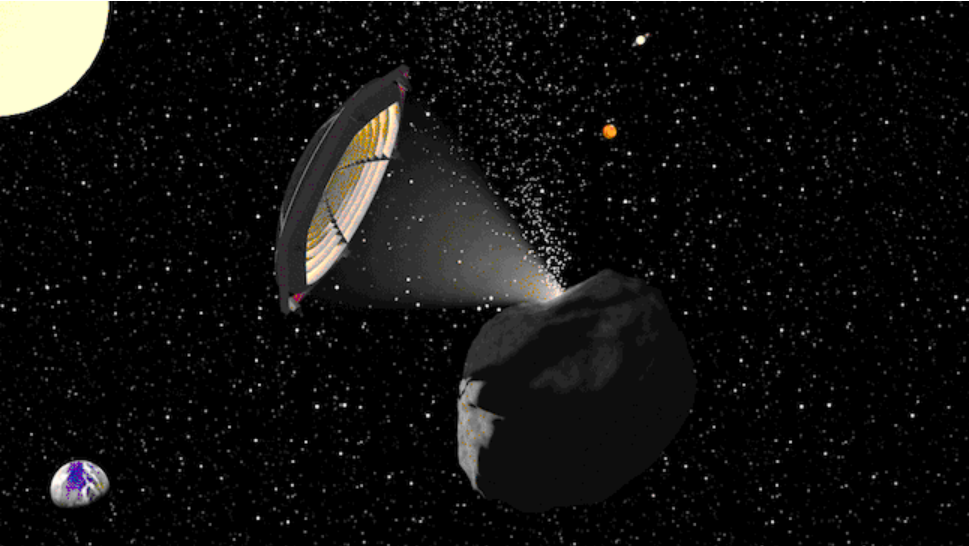}}\\
\subfigure[]{\includegraphics[width=8 cm]{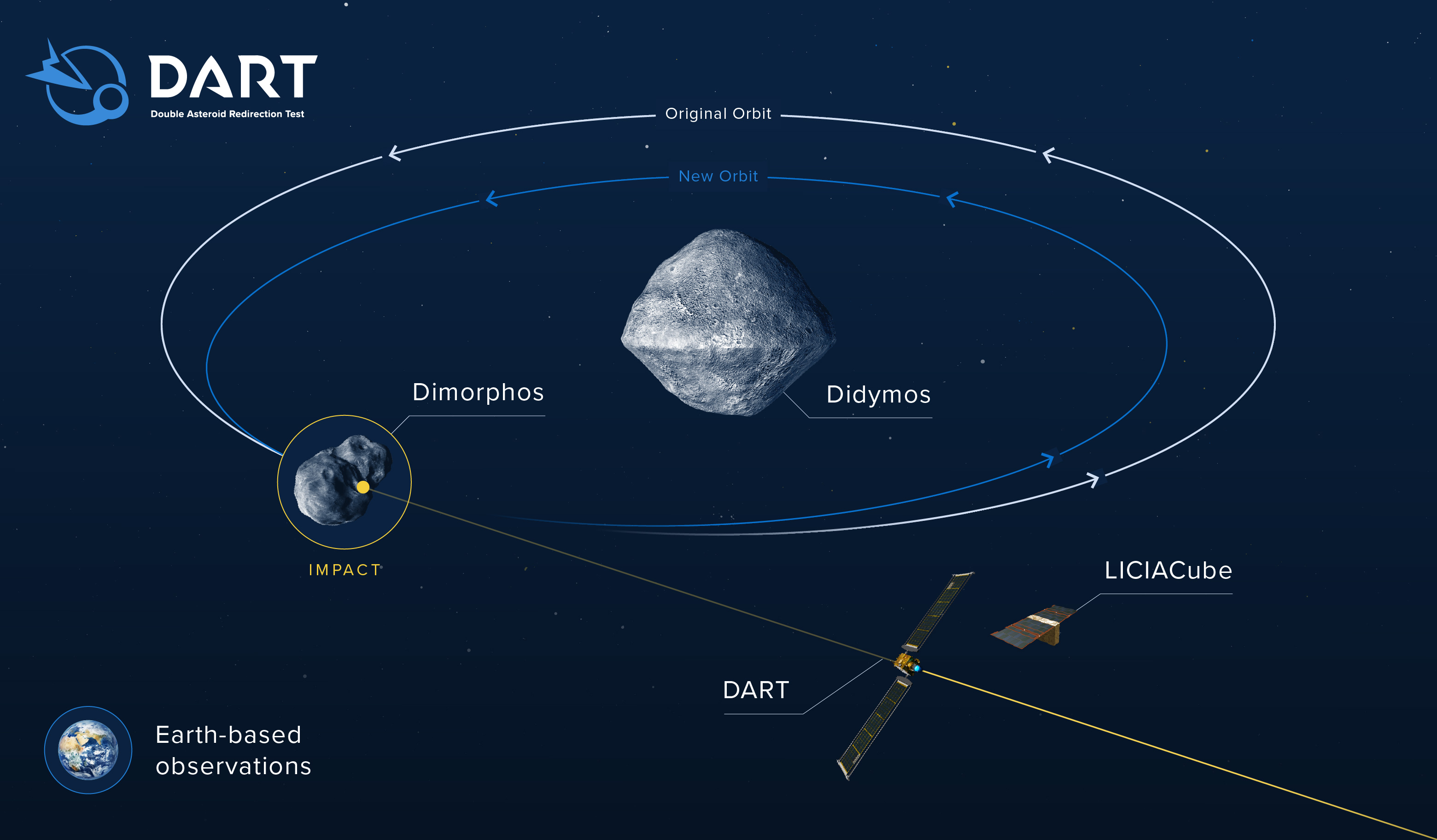}}\\
\subfigure[]{\includegraphics[width=8 cm]{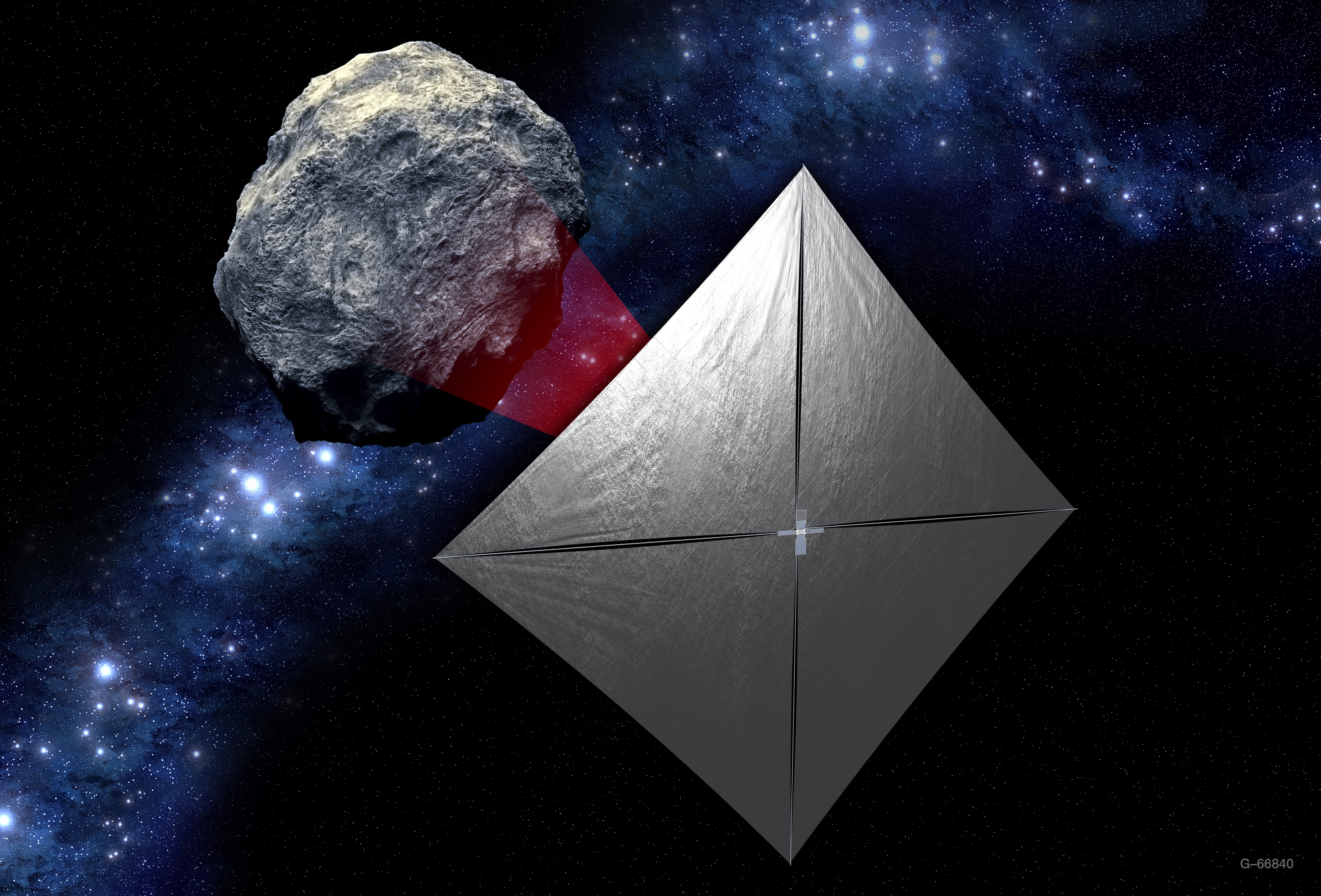}}
\caption{Illustration of various mitigation technologies of (\textbf{a}) gravity tractor (photo credit: NASA) and (\textbf{b}) focused solar energy (photo credit: Sergv22 CC BY-SA 4.0). (\textbf{c}) The deployment scheme of DART mission (	photo credit: NASA/Johns Hopkins APL). (\textbf{d}) NEA Scout with the solar sail deployed as it flies by its asteroid destination \cite{ref-NASA-Solar-Sail}. \label{Mitigation}}
\end{figure} 

\begin{figure*} %[H]
\centering
\subfigure[]{\includegraphics[width=12.5 cm]{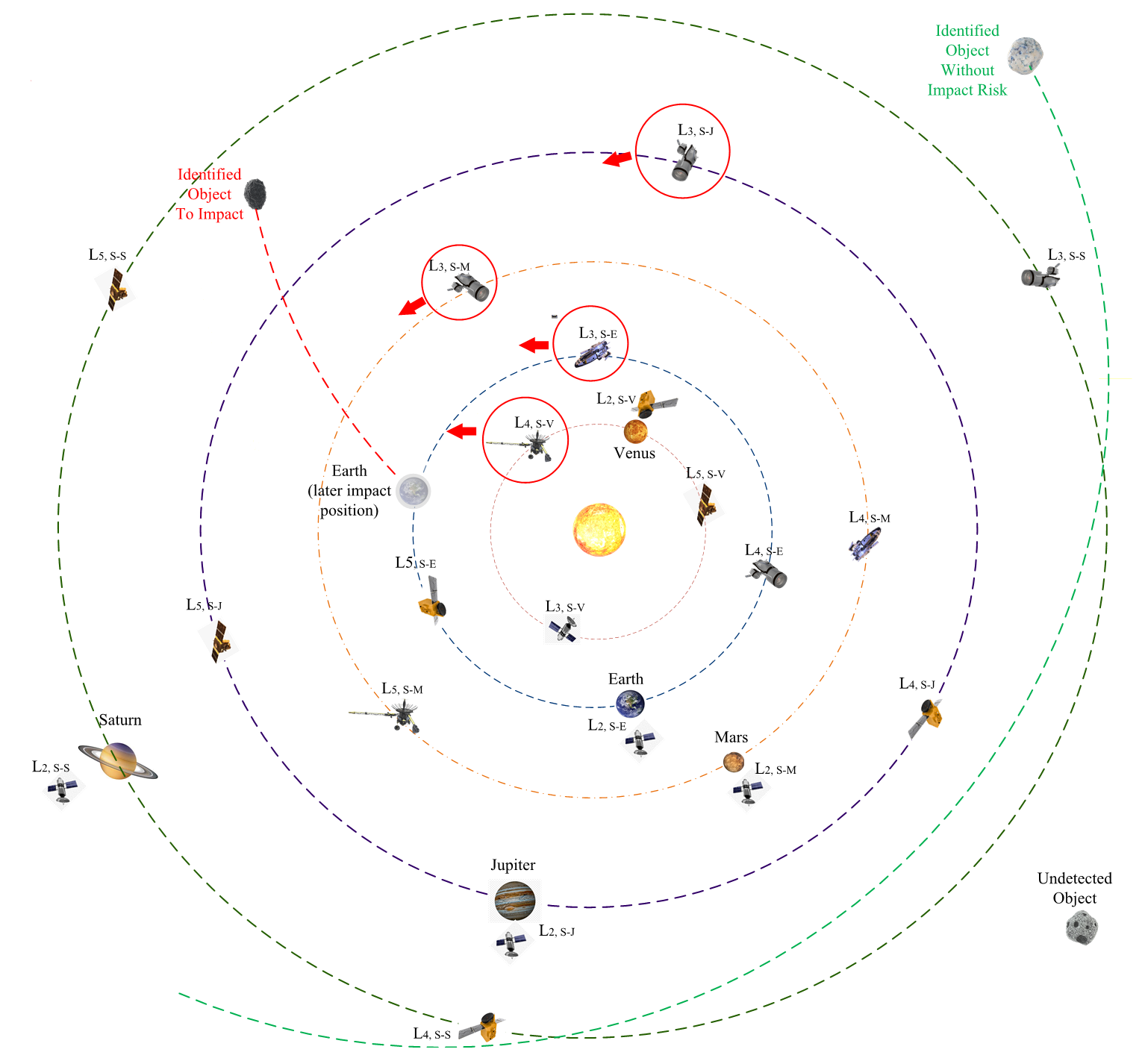}}\\
\subfigure[]{\includegraphics[width= 5 cm]{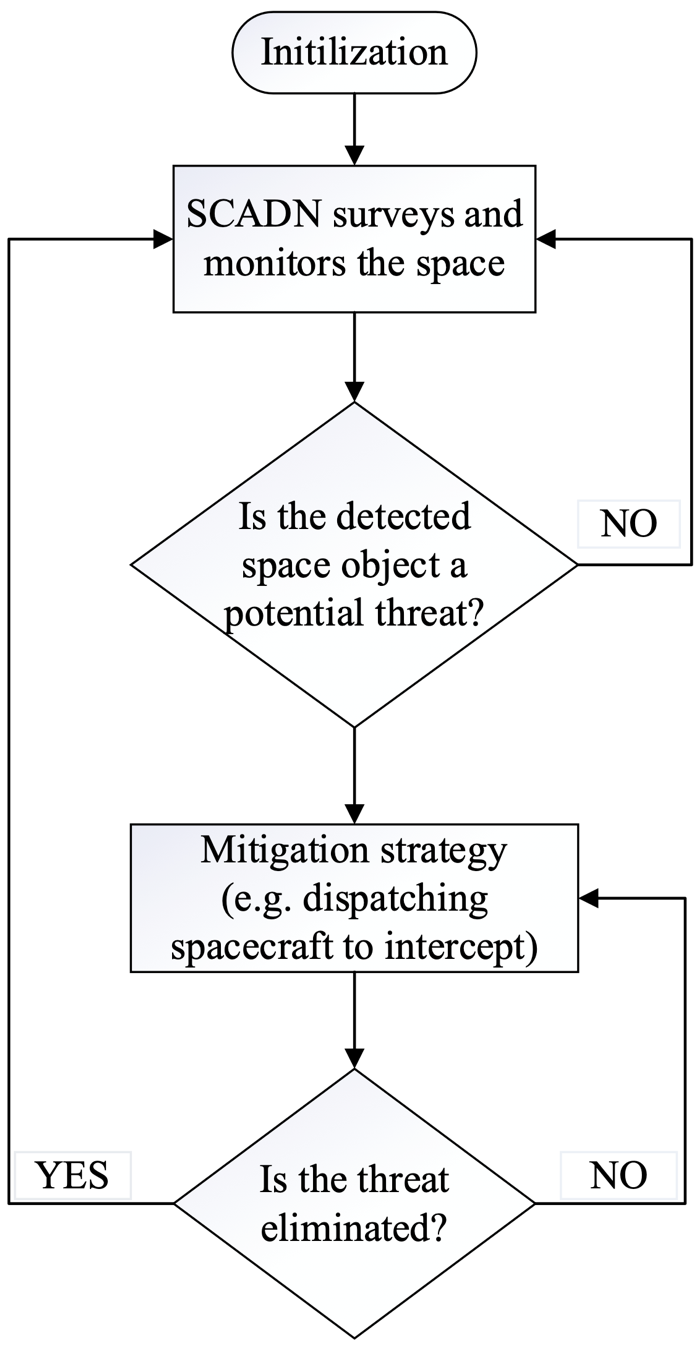}}
\caption{(\textbf{a}) Illustration of the SCADN framework which monitors the space, detects and intercepts the hazardous space objects (dimension of celestial bodies, space objects, and orbits are not scaled), and (\textbf{b}) a brief flow chart of the SCADN framework. \label{SCADN-INTERCEPT}}
\end{figure*}

\begin{table*} %[H]
\small
\caption{A summary of representative mitigation technologies.\label{table:mitigation}}
%%% \tablesize{} %% You can specify the fontsize here, e.g., \tablesize{\footnotesize}. If commented out \small will be used.
\newcommand{\tabincell}[2]{\begin{tabular}{@{}#1@{}}#2\end{tabular}}
\centering
\begin{threeparttable}
\begin{tabular}{|c|c|c|c|c|c|} \hline
%\toprule
 %\textbf{\tabincell{c}{Energy \\Source}}	&
  \textbf{\tabincell{c}{Energy \\Source}} & \textbf{Approach}	& \textbf{Strategy} & \textbf{\tabincell{c}{Key Technology\\and Features}} & \textbf{Example} &  \textbf{Comments} \\  \cline{1-6}
%\midrule
%\multirow[t]{4}{0.075\textwidth}{ } 
    &    \tabincell{c}{Stand-off \\approach} 	&  \tabincell{c}{Interception \\and\\trajectory-\\changing}  & \tabincell{c}{Denotation at\\20-m or greater\\stand-off height;\\10–100 times more\\ effective than the \\non-nuclear \\alternatives}    & \tabincell{c}{Project Icarus\\MIT Students \cite{ref-Day};\\ \\NASA's asteroid\\Interceptor \\ "Cradle spacecraft" \\in \cite{ref-Coppinger}\\  }  &  \tabincell{c}{In \cite{ref-Day}, a number of \\modified Satrun V \\rockets and creation \\of nuclear explosive\\devices in the 100-\\megaton energy range; \\In \cite{ref-Coppinger}, the conceptual \\spacecraft contains six \\B83 physics packages \\with each set for 1.2-\\megaton yield and\\and to be detonated \\over a 100-m height.}
    \\  \cline{2-6}

 	\tabincell{c}{Nuclear \\Explosive \\Device} &    \tabincell{c}{Surface and \\subsurface} & \tabincell{c}{Interception \\and\\trajectory-\\changing}  & \tabincell{c}{Creation of a conceptual \\Hypervelocity Asteroid \\Intercept Vehicle \\(HAIV),\\ which combines a \\kinetic impactor to \\create an initial crater\\ for a follow-up \\subsurface nuclear \\detonation within \\that initial crater\\ } 
 	
 	 &  \tabincell{c}{HAIV can cope with \\50-500-m diameter \\ objects when the \\time to Earth impact\\is less than one \\year \cite{ref-Nuking}; \\With a warning time \\ of 30 days, a 300-m\\wide asteroid can be \\neutralized by a single \\HAIV with less than 0.1\% \\of the PHO's mass \cite{ref-Wie} }
 	 
 	 & \tabincell{c}{HAIV can generate high \\degree of efficiency in \\the conversion of the \\nuclear energy \\that is released in \\the detonation into \\propulsion energy \\to the asteroid; \\ It may run an increased \\risk of fracturing the \\target NEO.}  \\ \cline{2-6}
	& \tabincell{c}{Comet \\Deflection}	& \tabincell{c}{Interception \\and\\vaporize or \\trajectory-\\changing} & \tabincell{c}{One-gigaton \\nuclear explosive \\device weighting\\25-30 tons, \\lifted on super-\\heavy rocket\\} 
	  
	  &   \tabincell{c}{Dr. Edward Teller \\proposed in 1995\\ Planetary Defense \\workshop \cite{ref-LLNL}\\}
	  &   \tabincell{c}{Instantly vaporize\\a 1-km asteroid \\or divert a 10-km one; \\It can cope with short-\\period comets escaping \\from Kuiper belt.\\} \\ \cline{1-6}
	  
	  \tabincell{c}{Kinetic \\Impact} & \tabincell{c}{Kinetic \\impactor\\deflection} & \tabincell{c}{Interception \\and \\ trajectory-\\changing} & \tabincell{c}{Sending spacecraft to \\a collision course \\to knock off \\the asteroid\\} &  \tabincell{c}{NEOShield-2 \\mission from ESA \cite{ref-NEOShield2}; \\Asteroid Impact and\\ Deflection Assessment \\(AIDA) missions of ESA/\\NASA, DART launched \\in Nov. 2021 \cite{ref-Double} \cite{ref-Planetary}}  & \tabincell{c}{The DART impact will \\occur in October 2022 \\and allow Earth-based \\telescopes and planetary \\radar to observe \\the event.}    \\ \cline{1-6}
 	  
 	  \tabincell{c}{Asteroid\\Gravity\\Tractor} &  \tabincell{c}{Apply a\\small but\\constant\\thrust}	& \tabincell{c}{Rendezvous\\with PHO\\and provide\\a small force} & \tabincell{c}{A massive unmanned  \\spacecraft hovering \\over an asteroid to \\gravitationally pull \\the asteroid into a \\non-threatening orbit} &  \tabincell{c}{Edward T. Lu and \\Stanley G. Love \\proposed \cite{ref-NASACongress} } &
 	   \tabincell{c}{The most expensive with\\the lowest technical \\readiness and many years \\to decades of duration \\might be required.}
 	  \\ \cline{1-6}
 	  
 	   \tabincell{c}{Focused\\Solar\\Energy} &  \tabincell{c}{Focus solar \\energy onto \\PHO's surface}	& \tabincell{c}{Remote\\station and\\rendezvous} & \tabincell{c}{Construction of remote \\station with large \\concave mirrors, \\concentration is scalable \\} &  \tabincell{c}{Proposed in \cite{ref-SolarNature}; \\Ring-array collector\\size is 0.5 PHO's\\diameter \cite{ref-Sunlight} } &
 	   \tabincell{c}{In \cite{ref-Sunlight}, 5,000 times the\\sunlight concentration,\\1,000-N thrusting effect,\\forming gas flow.}
 	   
 	  \\ \cline{1-6}
 	  
 	   \tabincell{c}{Asteroid\\Laser\\Ablation} &  \tabincell{c}{Focus laser\\onto PHO's \\surface}	& \tabincell{c}{Rendezvous\\and trajectory\\-changing} & \tabincell{c}{Concentrate laser \\energy to cause flash \\vaporization/ablation\\with reaction force} &  \tabincell{c}{First proposed in \cite{ref-LaserDefense}; \\Project DE-STAR, \\proposed \cite{ref-DE-STAR}} &
 	   \tabincell{c}{\cite{ref-DE-STAR} phased-array laser\\about 1 km squared,\\launched in increments\\assembled in space.}
 	  \\ \cline{1-6}
 	  
 	   	   \tabincell{c}{Ion\\Beam\\Shepherd} &
 	   	   \tabincell{c}{Pointing ion \\thruster at \\PHO's surface}	& \tabincell{c}{Interception\\and trajectory\\-changing} & \tabincell{c}{Use the
momentum \\transmitted by a low-\\divergence (< 15 deg) \\accelerated ion beam } &  \tabincell{c}{First proposed in \cite{ref-Ion Beam}} &
 	   \tabincell{c}{A 5-ton space debris can\\be deorbited in about 7\\months with IBS mass \\less than 300 kg.}
 	  \\ \cline{1-6}
 	 
\end{tabular}

%   \begin{tablenotes}
%      \footnotesize
        %\item[*] the European Space Agency and it consists of 22 countries.
%     \end{tablenotes}

\end{threeparttable}
\end{table*}

\begin{enumerate}
\item Ideally, a hybrid of PHO deflection schemes can be deployed to the spacecraft in each planet's Lagrange points. For example, in remote areas such as around Saturn, Uranus, or Neptune, the low-cost and slow-paced schemes (non-nuclear scheme) can work over an allowable time window that is usually large enough. For example, whenever a space object is detected and determined to be a treat to Earth or other humans' colonization by the SCADN, one or more spacecraft patrolling in the nearest proximity or being able to intercept the PHO on its collision course will be scheduled and coordinated to handle the mitigation. 

\item If the situation is so urgent that a fast response is needed immediately, the decision-making, resource-allocation, task-scheduling, and trajectory-planning can be fully autonomous, e.g., completely directed by AI/edge-computing, to overcome the large latency in space communication (due to extremely large space travel distance). The early detection and mitigation will result in lower cost and higher success rate of mitigation. The power supply can be a hybrid source of nuclear battery and solar panel, while the propulsive devices can be of various types, e.g., cold gas thruster, electrohydrodynamic thruster, electrodeless plasma thruster, electrostatic ion thruster, Hall effect thruster, magnetoplasmadynamic thruster, etc.  

\item An illustrative description of an exemplary application scenario is presented in Figure \ref{SCADN-INTERCEPT} (a) where a space object is detected and identified as having a high probability of impacting Earth within less than one year. The SCADN framework calculates its trajectory (in the red dash line), determines its mass and characteristics, and makes feasible mitigation strategies and plans. Eventually, the SCADN schedules the available spacecraft equipped with suitable mitigation technologies to intercept the object. During this procedure, the spacecraft perhaps need to change its orbit, accelerate or decelerate, which can lead to the consumption of power and thrust on board the spacecraft. In case one spacecraft fails a mitigation task, SCADN can call for multiple spacecraft simultaneously from different orbital locations of different planets to perform the interception. 

\item The SCADN framework is capable of computing and mobilizing all available resources within its framework to monitor any space object sensed, estimate the risk, and take the corresponding actions. Such a SCADN framework is supposed to be robust enough that it can still be functional to detect and mitigate the risk even when some spacecraft/survey stations within the framework cannot work properly or human operators are not available. A brief flow chart of the SCADN is given in Figure \ref{SCADN-INTERCEPT} (b), where the SCADN framework will utilize all available spacecraft to perform the interception until the impact risk is completely neutralized. Since all spacecraft being able to conduct the mitigation missions are continuously on alert in the space, the success rate is largely improved compared to any improvisational interception launch mission from Earth.        

\end{enumerate}

\subsection{Long-Term Universal Efforts}
Planning, developing, deploying, maintaining, and upgrading such a huge SCADN framework would require significant resources and efforts across many various entities. There will be expectations of critical collaborations among space agencies (e.g., NASA, ESA), governments, intergovernmental organizations (e.g., EU, ITU), research facilities, technical and professor associations (e.g., IEEE, AIAA), and technological corporations (e.g., SpaceX, Blue Origin), etc. 

Furthermore, the legislation of related laws and standards is another critical success factor. The famous astronomer and planetary scientist Dr. Carl Sagan, expressed concern about deflection technology in his book Pale Blue Dot \cite{ref-Sagan} where he noted that any method capable of deflecting impactors away from Earth could also be abused to divert non-threatening bodies toward the planet. Therefore, rigorous legislation and implementation are of the highest priority and importance during the top-down design of the proof-of-concept (PoC). For example, efficient public supervision and timely disclosure may help improve the eventual success of the SCADN framework. Also, developing and widely adopting advanced AI algorithms can decrease the unauthorized manipulation or mistakes of manual operations and enhance the cyber-security and overall performance of the SCADN framework. The authorization of upgrading the framework and core AI algorithms may request the highest-level granting from multiple entities to serve the common interest of all humankind best.     

In addition, several related international treaties should be taken into account when planning and developing the SCADN framework. The first is the Treaty on the Non-Proliferation of Nuclear Weapons, commonly known as the Non-Proliferation Treaty (NPT). It aims at preventing the spread of nuclear weapons and weapons technology to promote cooperation in using nuclear energy peacefully, with a further target of achieving nuclear disarmament, including general and complete disarmament \cite{ref-NPT}. The second one is The Outer Space Treaty (OST), formally the Treaty on Principles Governing the Activities of States in the Exploration and Use of Outer Space, including the Moon and Other Celestial Bodies. The OST is a multilateral treaty that forms the basis of international space law. In particular, it has included several key provisions such as prohibiting nuclear weapons in space and limiting the use of the Moon and all other celestial bodies to peaceful purposes \cite{ref-OST}. Consequently, there is a necessity for considerable investigation of the asteroid/comet mitigation technologies under both NPT and OST. The SCADN framework is likely to comply with these treaties since early detection can enable the use of slow and steady mitigation technologies.         

Moreover, fully deploying the SCADN framework may take significant time. Dividing the construction into several stages is thus reasonable and also feasible. For example, Phase I of the SCADN framework can target the coverage of the area reaching Jupiter. It is also noteworthy that several factors need to be considered as they can interrupt the progress to some extent. For instance, the global supply chain can be disrupted during the emergencies such as wars and pandemics \cite{ref-NatureGuan}.   

%%%%%%%%%%%%%%%%%%%%%%%%%%%%%%%%%%%%%%%%%%
\section{Conclusions}
In this paper, a comprehensive review of space exploration, space colonization, space threats, extinction events, space observatory, and space defense has been given, followed by a detailed presentation of related research and projects on milestones in human history. Furthermore, a summary and prediction are made to humanity's future evolution path: terraforming other celestial bodies for extra-terrestrialization. As a result, multi-planet-based communications and defense networks will be critically necessary, and a framework named Solar Communication and Defense Networks (SCADN) is proposed and analyzed. The distributed and intelligent features of the proposed SCADN framework can provide high reliability in coping with the potentially hazardous space objects in emergent situations and serve other purposes of space exploration and astronomical survey. Eventually, it is envisioned that founding such an enormous framework may require unprecedented resources and efforts across many nations and entities on a long-term basis for human living and prosperity. To utilize such a framework to serve the entire human race's common interest under international treaties, strict legislation, scientific implementation, and public supervision will play a crucial role.  

\vspace{6pt} 

\section{Acknowledgement}
The author would like to express sincere gratitude to the great minds and pioneers who have dedicated their efforts and even life to astronomy and space exploration.

\end{document}